\newcommand{\calcium}{Ca$^{2+}$}
\documentclass[]{article}
\usepackage{amssymb,amsfonts,amsthm,bm,graphicx,amsmath,float,color,longtable,needspace,parskip}
\restylefloat{figure}
\usepackage{tabulary,booktabs}%,chemformula}
\usepackage[ansinew]{inputenc}
\usepackage{lscape}
\usepackage[backend=bibtex, style=nature]{biblatex}
\usepackage{hyperref}
\renewcommand{\baselinestretch}{1.15}
\renewcommand{\baselinestretch}{1.5} 
\begin{document}

\title{Paradoxical signaling regulates structural plasticity in dendritic spines}

\author{Padmini Rangamani$^{1*}$,  Michael G. Levy$^2$,  Shahid M. Khan$^3$,  and George Oster$^4$
\\[0.1in]
$^1$ Department of Mechanical and Aerospace Engineering,\\
 University of California San Diego, La Jolla CA 92093\\
$^2$ Biophysics Graduate Program, University of California Berkeley, \\
Berkeley CA 94720\\
$^3$ Molecular Biology Consortium, Lawrence Berkeley National Laboratory, \\
Berkeley CA 94720\\
$^4$ Department of Molecular and Cell Biology, University of California Berkeley, \\
Berkeley CA 94720.\\
$^*$ To whom correspondence should be addressed: padmini.rangamani@eng.ucsd.edu.
}
%\date{}
\maketitle

\abstract{Transient spine enlargement (3-5 min timescale) is an important event associated with the structural plasticity of dendritic spines. Many of the molecular mechanisms associated with transient spine enlargement have been identified experimentally. Here, we use a systems biology approach to construct a mathematical model of biochemical signaling and actin-mediated transient spine expansion in response to calcium-influx due to NMDA receptor activation. We have identified that a key feature of this signaling network is the paradoxical signaling loop. Paradoxical components act bifunctionally in signaling networks and their role is to control both the activation and inhibition of a desired response function (protein activity or spine volume). Using ordinary differential equation (ODE)-based modeling, we show that the dynamics of different regulators of transient spine expansion including CaMKII, RhoA, and Cdc42  and the spine volume can be described using paradoxical signaling loops. Our model is able to capture the experimentally observed dynamics of transient spine volume. Furthermore, we show that actin remodeling events provide a robustness to spine volume dynamics. We also generate experimentally testable predictions about the role of different components and parameters of the network on spine dynamics.}

\section*{Introduction}
The ability of the brain to encode and store information depends on the plastic nature of the individual synapses. The increase and decrease in synaptic strength, mediated through the structural plasticity of the spine, is important for learning, memory and cognitive function \cite{Colgan2014,lisman2002,nimchinsky2002}. Dendritic spines are small structures that contain the synapse. They come in a variety of shapes (stubby, thin, or mushroom-shaped), and a wide-range of sizes that protrude from the dendrite \cite{gray1959,harris1994}. These are the regions where the postsynaptic biochemical machinery responds to the neurotransmitters \cite{Colgan2014}. Spines are dynamic structures, changing in size, shape, and number during development and aging \cite{yang2009,shibata2015,hedrick2014,nimchinsky2002}.  

\begin{figure*}
\centerline{\includegraphics[width=\textwidth]{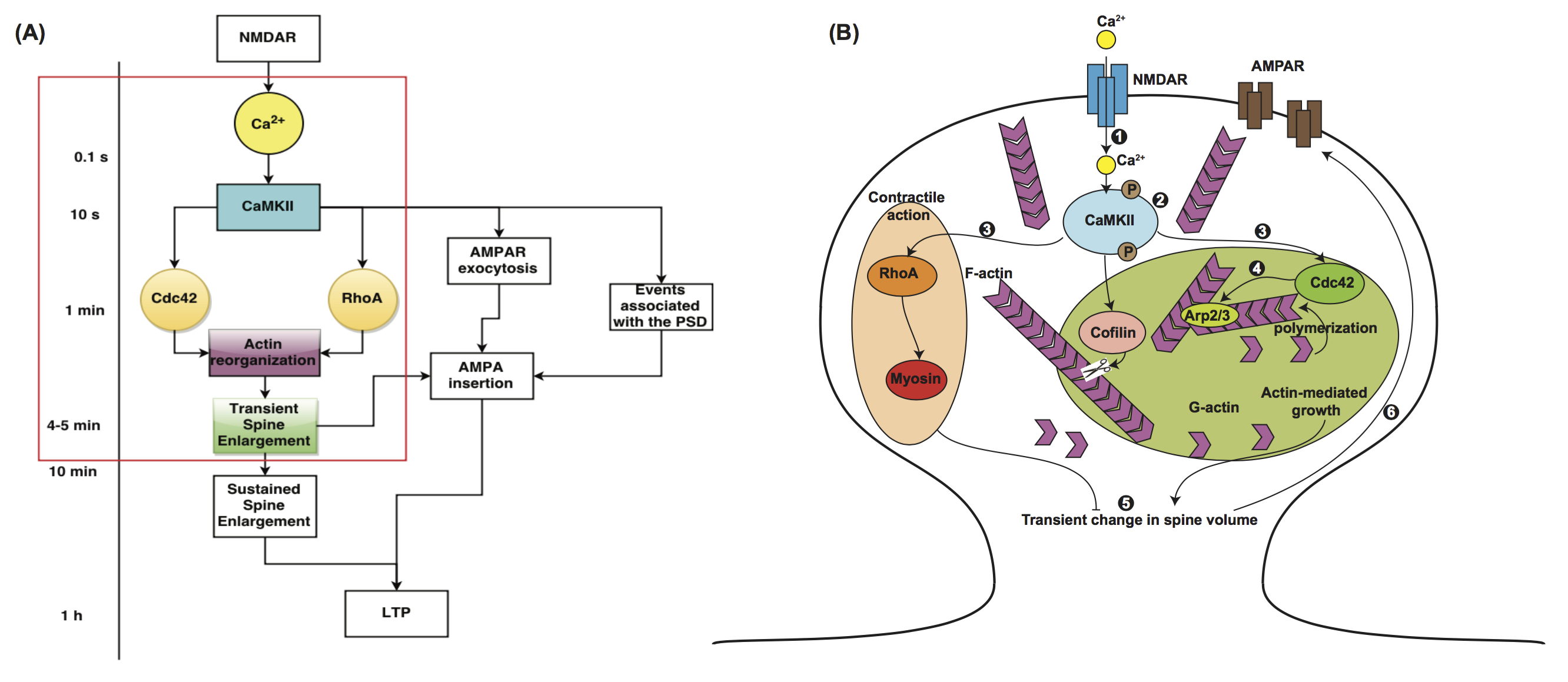}}
\caption{\calcium-CaMKII regulation of short-term and long-term events in LTP. (A) The time scales associated with the events leading up to LTP are shown here. NMDA receptor activation leads to \calcium release at the millisecond time scale. CaMKII activation by \calcium is rapid and occurs within tens of seconds. The small Rho-GTPases, Cdc42 and RhoA are activated within a minute or so and lead to actin reorganization events resulting in transient enlargement of the spine in 4-5 min. Other events include AMPA receptor exocytosis and insertion in the plasma membrane of the synapse, reorganization of the post-synaptic density (PSD), and long-term potentiation that takes place at the time scale of an hour. Our focus is on the events leading up to the  transient spine enlargement, as highlighted by the red box. This figure was adapted from \cite{Murakoshi2012}. (B) The event associated with transient spine enlargement are shown in the schematic. The numbers correspond to the following events. (1)  Binding of NMDA to NMDA receptor results in a \calcium influx. (2) \calcium influx induces a rapid and transient activation of \calcium/Calmodulin-dependent protein kinase II (CaMKII). (3) CaMKII activation is followed by local and persistent activation of Rho GTPases in the spine. (4)  Rapid remodeling of the actin cytoskeleton takes place in response to CaMKII and Rho GTPase activation. (5)These signaling cascades result in a transient increase in spine volume. (6) Synaptic strength is enhanced and made long-lasting by the insertion of functional $\alpha$-amino-3-hydroxy-5-methyl-4isoxazolepropionic acid (AMPA) receptors in the spine membrane.}
\label{fig:figure0}
\end{figure*}

Recent advances in imaging techniques have allowed neuroscientists to study the dynamics of the change in spine volume and identify the role of different molecular components in mediating the structural plasticity of the spine. One way to induce long-term potentiation (LTP) is through  N-methyl-D-aspartate (NMDA)-receptor mediated signaling. The resultant transient and sustained changes in the size of the dendritic spine are important for  long-term potentiation.  LTP is a complex phenomenon, but the events underlying this phenomenon can be summarized as a multiple time scale process as shown in Figure \ref{fig:figure0}A. In this study, we focus on the events occurring in the 3-5 min time scale, shown in the red box in Figure \ref{fig:figure0}A. The early events in LTP can be summarized as follows and schematically represented in (Figure~\ref{fig:figure0} B).

The initial stimulus is the NMDA receptor inward calcium current in response to electrical and chemical stimulation. The NMDA receptor inhibitor APS blocks both current and LTP.  Influx of calcium is tightly coupled to spine expansion a minute or so later followed by partial compaction over the next 3-4 minutes \cite{Bosch2014}. Remodelling of the actin spine cytoskeleton coupled to spine expansion is triggered by CaMKII, which is activated by calcium calmodulin. Free G-actin is recruited for polymerization \cite{Hoffman2013}. This activation has two consequences. One, activated CaMKII dissociates from F-actin and associates with the post-synaptic density (PSD) \cite{okamoto2009,Kim2015}. There is an influx of CaMKII, cofilin, debrin and Arp 2/3 from dendrites to the stimulated spine, presumably due to F-actin sites vacated by the activated CaMKII.  Two, activated CaMKII phosphorylates multiple targets. These include the small GTPases Rho and Cdc42, that in turn, initiate phosphorylation cascades. The effects of Rho are localized to the stimulated spine, whereas Cdc42 phosphorylates targets in adjacent spines to facilitate later stages of LTP \cite{Murakoshi2012}.  Phosphorylated CaMKII has a limited half-life, since calcium calmodulin also activates PP1 phosphatase that dephosphorylates CaMKII, albeit with a slower timescale \cite{Pi2008}.
Myosin IIb isoforms are involved in maintenance of spine morphology as established by studies with the myosin IIb inhibitor blebbistation and siRNA \cite{Koskinen2014,ryu2006}. Non-muscle and sarcomeric myosin IIb isoforms play distinct roles in early LTP \cite{kneussel2013,Rubio2011}. The non-muscle isoform localizes at the spine base where it might stabilize the actin cytoskeleton. The sarcomeric isoform associates with SynGap1 at the post synaptic density (PSD). Its contractile activity triggered by myosin light chain kinase (MLCK) could drive spine head compaction that produces the subsequent decrease in spine volume. The volume decrease does not reach pre-stimulated levels. One attenuating factor could be partial disassembly of the compacted F-actin cytoskeleton. 

Experiments have revealed that different molecules and their complex interactions regulate synaptic plasticity \cite{coba2009,collins2005}. While the temporal response of the spine volume change and the associated molecular events is well-documented \cite{Murakoshi2012,Murakoshi2011,Lee2009}, conceptual models cannot adequately explain the dynamics of these molecular processes. Computational models, on the other hand, can provide insight into the nonlinear molecular interactions, identify network motifs, and also generate experimentally testable predictions.  Dynamical modeling using ODEs approaches have been used to test hypotheses \textit{in silico}, generate time courses, and identify emergent properties that would be hard to investigate experimentally. This approach has been used in neuroscience successfully to study different aspects of structural plasticity \cite{bhalla1999,Neves2008,Pi2008}. In this study, using computational models, we seek to answer the following questions -- (a) Is there a network motif that regulates the dynamics of CaMKII, RhoA, and Cdc42 in response to \calcium-influx? (b) How does biochemical signaling interface with actin remodeling to control the transient expansion of the spine? In what follows, we outline the construction of the signaling network and the associated mathematical model, compare model outcomes against experimentally observed dynamics of various components, and finally generate experimentally testable predictions.

The dynamics of CaMKII, RhoGTPases, and spine volume in the 3-5 minute time scale exhibit a similar profile but with different time scales \cite{Kim2014,Murakoshi2011}. These dynamics can be explained by biexponential functions. One network motif that gives rise to biexponential functions is a paradoxical signaling loop. Paradoxical signaling is a network motif where the same stimulus controls the response by regulating both its activation and inhibition. Paradoxical signaling has been identified in different contexts; examples include cellular homeostasis, and cell population control \cite{Alon2007,Hart2012,Hart2013}. In these studies, the simple idea that the same component can both activate and inhibit a response resulting in robustness was applied to many different systems. Here we consider the same design principle for the dynamics of the dendritic spine and apply it to the core signaling network that regulates spine dynamics.

\section*{Model Development}
\subsection*{Modular construction of the reaction network}
 Our first step was to analyze the experimental time course of CaMKII, Rho, Cdc42, and spine volume to predict a network motif based on the principles of chemical reaction engineering. Then, a biochemical signaling network was constructed  \cite{azeloglu2015,azeloglu2015b}.
CaMKII \cite{Lee2009}, small RhoGTPases \cite{Murakoshi2011}, and actin  and related components \cite{Okamoto2007,kim2013,Kim2014,okamoto2009,Calabrese2014,kneussel2013} play important roles in the transient volume change in the spine. Based on these studies, we identified the main components that regulate spine dynamics as CaMKII, Arp2/3, cofilin, Rho, actin, and myosin Figure \ref{fig:figure1}A. In the model presented below, myosin II will refer to the sarcomeric myosin IIb isoform \cite{Blanchoin2014}. We note here that while there are many more biochemical components involved in the signaling network in the single spine, our choice of components was based on experimentally measured dynamics. The expanded network is shown in Figure \ref{fig:figure1}B and the complete table of reactions for each module is given in the Supplementary Material (Tables \ref{table:CaMKII}--\ref{table:Rho}). We assumed that the components were present in large-enough quantities that concentrations could be used to represent the amounts of the different molecular species. This assumption allowed us to generate a deterministic dynamical model. We also assumed that the spine is a well-mixed compartment so that we could follow the temporal evolution of the concentrations of the different components. Each interaction was modeled as a chemical reaction either using mass-action kinetics for binding-unbinding reactions, and Michaelis-Menten kinetics for enzyme-catalyzed reactions \cite{bhalla1999,rangamani2008}.  The network of interactions was constructed using the Virtual Cell modeling platform (http://www.nrcam.uchc.edu). 
 
In response to the \calcium-CaMKII activation, the F- and G-actin released from CaMKII are released and free to undergo actin polymerization events \cite{okamoto2009,Okamoto2004,Okamoto2007}. Since Cdc42 and Rho are activated downstream of CaMKII, we hypothesize that these lead to Arp2/3, cofilin, and myosin activation. This module is built based on the literature evidence of the role of Arp2/3, cofilin and myosin in the dendritic spine. We linked the Cdc42 activation to Arp2/3 and actin barbed ends production using a slightly modified version of the model presented by Tania \textit{et al.} \cite{Tania2013}. In our model, we included both G- and F-actin. Once the CaMKII-bound actin is released, it undergoes remodeling to generate a large number of barbed ends catalyzed by Arp2/3 and cofilin \cite{Tania2013,kim2013}.

\subsection*{Characteristics of the signaling cascade}
In order to characterize the dynamics of the different protein activities and the volume change of the spine, we use the characteristics of the time-concentration or time-radius curve \cite{Heinrich2002}. We use three measures of the curve to characterize the model output. (1) Time to peak signal -- this is the time point at which maximum signal is achieved. (2) The area under the curve gives the total signal activated over the time of observation and for the $i^{th}$ species is given by $I_i$ in Eq. \ref{eq:characteristics}. (3) The signaling time is given by $\tau_i$ and gives the average duration of signal and is also defined in Eq. \ref{eq:characteristics}. These different values can be interpreted as the statistical measures of a time course in a signaling network (Table \ref{table:characteristics}).
\begin{equation}
I_i=\int_0^{\infty} X_i(t)dt, \quad \text{and} \quad \tau_i=\frac{\int_0^{\infty} tX_i(t)dt}{I_i}.
\label{eq:characteristics}
\end{equation}

\subsection*{Comparison with experimental data}
The experimental data was extracted from Figure 1D of Murakoshi \textit{et al.} \cite{Murakoshi2012} using the digitize package in the statistical software `R'. Complete details of the digitize package and how to use it are provided in \cite{poisot2011}. Since the experimental data was normalized, we also normalized the simulation data to the maximum. Experimental and simulation data was compared for goodness of fit using root-mean-squared error.  A root-mean-squared error of less than 0.30 was considered to represent a good fit.

\subsection*{Dynamic parametric sensitivity analysis}
We conducted a local parametric sensitivity analysis of the model to identify the set of parameters and initial concentrations that govern model robustness. The log sensitivity coefficient of the concentration of the $i^{th}$ species $C_i$, with respect to parameter $k_j$ is given by \cite{varma2005,ingalls2003}

\begin{equation}
S_{i,j}=\frac{\partial \ln{C_i}}{\partial \ln{k_j}}
\label{eq:sensitivity}
\end{equation}

Since we are dealing with a dynamical system and not steady state behavior, we used the Virtual cell software to calculate the change in log sensitivity over time ($\frac{dS_{i,j}}{dt}$). The resulting time course gives us information about the time dependence of parametric sensitivity coefficients for the system. The variable of interest, $C_i$ is said to be robust with respect to a parameter $k_j$ if the log sensitivity is of the order 1 \cite{varma2005}. We conducted dynamic sensitivity analysis not only for all the kinetic parameters in the system but also for the initial concentrations of the various species in the model.

\section*{Results}
\subsection*{Paradoxical signaling}
A summary of the key experimental observations for transient activity in the dendritic spine is given in Figure 1D of \cite{Murakoshi2012}. Two qualitative features of these dynamics are important -- (1) they can  be fit to bi-exponential functions, where one exponent controls the time scale of activation and the other controls the time scale of deactivation (Figure \ref{fig:figures1}A). (2) A series of nested biexponential functions can be tuned so as to move the peak activation time of the different components relative to one another (Figure \ref{fig:figures1}B). 
Consider the differential equation, 
\begin{equation}
\frac{df}{dt}=\frac{ab}{a-b}(\underbrace{ae^{-at}}_{activation}-\underbrace{be^{-bt}}_{inhibition}),
\end{equation}
where $a$ and $b$ are constants. From this form, it is easy to see that the activation of $f$ is governed by $e^{-at}$ and the inhibition is governed by $e^{-bt}$. 
The solution to this equation for constant $a$ and $b$ is given by the biexponential function 
\begin{equation}
f(t)=\frac{ab}{a-b}(e^{-bt}-e^{-at}).
\label{eq:biexponential}
\end{equation}
When $a=b$, the function is given by $f(t)=a^2te^{-at}$. In general, the coefficients $a$ and $b$ need not be constant but can be functions of time or other component concentrations. This function produces the profile observed for the dynamics of CaMKII, Rho, Cdc42, and volume for different values of $a$ and $b$ (Figure \ref{fig:figures1}A). 

The dynamics of a simple activation-inhibition scheme driven by the same stimulus is shown in Figure \ref{fig:figure2}A. In response to the calcium input (stimulus), concentrations of both the activators and inhibitors of spine volume are increased. The net change in spine volume is then a balance between the effect of the activator and inhibitor. The structure of the network in Figure \ref{fig:figure2}A suggests that the same stimulus $S(t)$ both activates and inhibits the response $R(t)$, by controlling the level of $A(t)$ and $I(t)$ simultaneously. In this study, we propose that \calcium mediated activation of CaMKII results in a paradoxical signaling structure where CaMKII regulates both the expansion and contraction of the spine volume by regulating actin dynamics through intermediaries such as Rho, Cdc42, and other actin related proteins. This indicates that a simple paradoxical signaling may carry the kinetic information necessary to represent the time-dependent volume changes in the dendritic spine. 

\subsection*{Multi-tier model of paradoxical signaling}
Biological signaling networks are quite complex and have multiple interconnected motifs. These interconnections play an important role in regulating the dynamics of different components in a signaling cascade. One way to alter the time scale of the response in the one-tier model of paradoxical signaling is to change the kinetic parameters of the model shown in Figure \ref{fig:figure2}B. Another way to change the time scale of the process is to add multiple tiers to slow down the flow of information from the stimulus to the response. 

While a simple model of paradoxical signaling provides us insight into the basic structure of the network, biological signaling often contains many proteins working in concert to regulate a process. By adding a series of activators (activator 1 and 2 to mimic the actin-related proteins and actin barbed end generation) and inhibitors (inhibitor 1 and 2 to represent Rho and myosin) (Figure \ref{fig:figure2}C), we can control the response dynamics (spine volume) (Figure \ref{fig:figure2}D). The full set of associated ordinary differential equations and parameters, along with analytical solutions are given in the supplemental material (Table \ref{table:multitier}). 

The modules in Figure \ref{fig:figure1}A can be further expanded to construct the biochemical reaction network in Figure \ref{fig:figure1}B based on experimental observations and the corresponding details are presented in the supplementary material. An emergent property of this network is that despite its biochemical complexity, each module within this network (CaMKII, Arp2/3, etc.) also exhibits the same structure as Figure \ref{fig:figure2}A, B. As a result, we suggest that a series of nested paradoxical signaling loops control the response of the spine volume to the calcium influx and the nesting of these loops serves to control the time scale of the response to a calcium influx (Figure \ref{fig:figure2}C, D). 

One of the key differences between the modular representation in Figure \ref{fig:figure1}A and Figure \ref{fig:figure1}B is the cross-talk between components that participate in the activation and inhibition loop. For example, Rho activates ROCK (Rho kinase), which then regulates cofilin activity through LIM kinase and MLCK activity. Another example of the shared GAP for Rho and Cdc42. These interactions, which cannot be represented in the simple modular structure of Figure \ref{fig:figure1}A, are important features of the complexity of biological signal transduction in the spine. 
 
 The complete table of reactions for each module, along with kinetic parameters, initial concentrations, model assumptions, and references in provided in the Supplementary Material. 

\subsection*{CaMKII activation is regulated by autophosphorylation and an ultrasensitive phosphatase cascade}
In response to glutamate binding to NMDA receptors, a \calcium-pulse is released in the postsynaptic dendrite \cite{Lee2009}. This \calcium  binds with Calmodulin, a calcium-binding messenger protein. Calcium-calmodulin activates CaMKII by cooperative binding. CaMKII can undergo autophosphorylation, leading to a higher activity level \cite{lisman2001,lisman2002}. CaMKII is dephosphorylated by protein-phosphatase 1 (PP1) through a series of phosphatases. In this module, the stimulus is calcium,  activators for CaMKII phosphorylation are calcium-calmodulin and CaMKII itself, and the inhibitors are PP1. PP1 can also undergo auto-dephosphorylation \cite{lisman2002}. The module structure is shown in Figure \ref{fig:figure3}A. Model simulations show that CaMKII dynamics are very sensitive to the concentration of PP1 and the autophosphorylation of CaMKII, coupled with the ultrasensitive phosphatase cascade and PP1 autodephosphorylation results in a bistable response (Figure \ref{fig:figures2}) \cite{genoux2002}, similar to the observations made by Zhabotinsky \cite{Zhabotinsky2000}. By changing the amount of PP1 in the system, we obtain either sustained activation of CaMKII, which is undesirable, or a transient activation, which is the desired response (Figure \ref{fig:figure3}(B)). Our model matched the experimentally observed time course of CaMKII activation when the concentration of PP1 was $0.36$ $\mu M$ (Figure \ref{fig:figure3} C).  

\subsection*{Rho and Cdc42 are regulated by CaMKII-mediated paradoxical signaling}
Rho and Cdc42 are small GTPases, whose activity is regulated by guanine nucleotide exchange factors (GEFs) to convert them from the -GDP bound form to the -GTP bound form and GTPase activating proteins (GAPs) to hydrolyze the bound GTP to GDP and subsequently inactivating the G-protein. This GTPase switch has been well-studied in different systems \cite{bourne1990,cherfils2011}. In the case of spine volume change, we modeled the regulation of both the GEF and GAP activity as being controlled by the CaMKII activity \cite{okamoto2009}. As a result, the paradoxical signaling network loop that controls Rho and Cdc42 activity is the one shown in Figure \ref{fig:figure4}A,B. The time course of normalized activation of Rho and Cdc42 matches with the experimental data very well (Figure \ref{fig:figure4}C), suggesting that the temporal control of spine dynamics depend on the network structure. 

\subsection*{Spine volume change in response to calcium influx as a nested paradoxical signaling network}
Transient change in the spine volume is an important part of LTP \cite{Bosch2014}.  The relationship between pushing barbed ends, $B_p$, and the membrane velocity, $V_{mb}$, has been derived in \cite{Lacayo2007,Tania2013}, and is used here to calculate the radius of the growing spine in response to the actin remodeling events. This relationship is given as
\begin{equation}
V_{mb}=V_0\frac{Bp}{Bp+\phi \exp(\omega/Bp)}
\label{eq:vmb}
\end{equation}
Here, we used the values of $\phi=10/\mu m$, $V_0=0.07 \mu m/s$,and $\omega=50$ per $\mu m$ \cite{Tania2013}. $\phi$ is the geometric parameter used in computing membrane protrusion rate and $\omega$ is the physical parameter describing the membrane resistance \cite{Tania2013,Lacayo2007}. This density-velocity relationship has a biphasic behavior -- for a small number of barbed ends, the membrane resistance limits the velocity, explained as `decoherent' regime in \cite{Lacayo2007} and for large barbed end density, or the `coherent' regime, the protrusion rate is not sensitive to the number of barbed ends. 

 We also assume that the Rho activation leads to the activation of the Rho kinase ROCK, and myosin phosphorylation \cite{kaneko2012}. We then propose that the increase in spine size is proportional to the increase in actin barbed ends (similar to the leading edge pushing velocity), and the decrease is proportional to the amount of phosphorylated myosin. 
The rate of change of spine size is given by the equation 
\begin{equation}
	\frac{dR}{dt}=\underbrace{V_{mb}}_{\text{actin dependent growth velocity}}-\underbrace{k_{shrink}[MLC^*]R}_{\text{myosin-mediated contractility}}
	\label{eq:spine_volume}
\end{equation}

 Even though the spine volume is not a concentration of chemical species, the structure of the Eq. \ref{eq:spine_volume} indicates the paradoxical nature of the dynamics. Using this model, we have identified a simple and intuitive network structure that controls the dynamics of the transient dendritic spine volume change. The large concentration of actin in the spine ensures that the number of barbed ends generated is large and the system is in its `coherent' state \cite{Lacayo2007}. 
 
The normalized change in spine radius with time is compared for model and experiment in Figure \ref{fig:figure5}. The paradoxical structure shown in Figure \ref{fig:figure5}A leads to the actin barbed-end based pushing velocity and myosin mediated pulling rates as shown in Figure \ref{fig:figure5}B. The actin mediated pushing occurs earlier than the myosin mediated pulling, resulting in spine volume change as shown in Figure \ref{fig:figure5}C. The root-mean-squared error between the experimental data and the simulation results was $0.04$, indicating that the model is able to capture experimentally observed dynamics quite well.

\subsection*{Actin nucleation and severing}
Actin barbed ends can be generated both by branching (Arp2/3-mediated) and severing (cofilin-mediated). To identify the contributions of these two components to barbed-end generation, we varied the concentrations of Arp2/3 and cofilin and studied their effect on pushing barbed end generation. For a fixed Arp2/3 concentration of $2 \mu M$, increasing cofilin concentrations leads to a sharp increase in the severing rate, $f_{sev}$ (Figure \ref{fig:nuc_sev}A). As a result, the generation of pushing barbed ends depends strongly on cofilin concentration (Figure \ref{fig:nuc_sev}B). On the other hand, increasing Arp2/3 concentration, with fixed cofilin concentration of $2 \mu M$, increases the nucleation rate $f_{nuc}$ only to a small extent (Figure \ref{fig:nuc_sev}C) and therefore there is no appreciable difference in pushing barbed end production for increasing Arp2/3 concentration (Figure \ref{fig:nuc_sev}D). These results suggest that when cofilin is present in sufficient quantities, it dominates the pushing barbed end generation, driving the system to the `coherent' regime, and Arp2/3 concentration plays a small role in controlling barbed end generation. On the other hand, when cofilin concentrations are small, Arp2/3 contribution to barbed end generation will be significant.

\subsection*{Sensitivity to initial conditions}
Using dynamic sensitivity analysis, we analyzed the sensitivity of CaMKII, Cdc42, Rho, and spine radius dynamics to the initial concentrations of the different components (Table \ref{table:initial_index}, Figure \ref{fig:initial_conditions}). In Figure \ref{fig:initial_conditions}, white is no sensitivity, and the pink-purple shades show increasing sensitivity. As listed in Table \ref{table:initial_index}, the biochemical regulation of CaMKII through the network of components shown in Figure \ref{fig:figure1} means that it is sensitive to many initial concentrations, at both early and later time stages. Similarly, Cdc42 and Rho show sensitivity to many initial conditions; while Cdc42 shows sensitivity at both early and later times, Rho shows sensitivity to initial conditions listed in Table \ref{table:initial_index} at later times.  However, the radius of the spine behaves in a robust manner and demonstrates sensitivity to fewer components. Predominant among those are the phosphatases that regulate CaMKII dynamics, and therefore actin binding to inactive CaMKII, and ROCK, which is responsible for myosin activity. These results suggest that the spine radius is quite robust to changes in initial concentrations of the different species even though the biochemical regulation of upstream components shows a larger sensitivity. 

\subsection*{Sensitivity to kinetic parameters}
We conducted sensitivity analysis to identify the kinetic parameters that the model is most sensitive to (Table \ref{table:kinetic_index}, Figure \ref{fig:kinetic_parameters}). We found that the model is mostly robust to small changes in many parameters. CaMKII was mostly sensitive to parameters that govern the autophosphorylation activity, while Cdc42 dynamics was mostly robust to all parameters. Rho activity was sensitive to parameters associated with CaMKII activation and inactivation. Most interestingly, the radius of the spine showed sensitivity only to the parameters that govern the actin dynamics, $V_0$ and $k_{cap}$. These results indicate that actin dynamics and therefore the spine radius change are robust to changes in the biochemical events preceding the actin remodeling, as long as sufficient barbed ends can be generated. 

\section*{Discussion}
The role of dendritic spine dynamics in development and learning has been well-established for a long time. The presence of contractile actin in these dynamic structures was established in the 1970s and provided a link between motility, spine dynamics, and memory and learning. The smallest length and time scale changes to spine shape and size are now known to impact higher-order synaptic, neural, and brain functions, including many pathologies and age-related mental disorders. Advances in microscopy have allowed us to study spine dynamics in different experimental settings from \textit{in vitro} cell culture to \textit{in vivo} animal models. However, there remain unanswered questions on how transient changes to spine volume are regulated by the molecular components. In this work, we propose that the transient dynamics to the spine volume are regulated by a simple paradoxical signaling module.

The dynamics of CaMKII, Cdc42, Rho, and spine volume all follow similar dynamics but at different time scales (Figures \ref{fig:figure3}, \ref{fig:figure4}, \ref{fig:figure5}). Since each of these curves can be fit to a biexponential function with different time scales of activation and inhibition, we wondered if these different phenomena can be explained by a simple regulatory structure. We identified that the biexponential function could be a result of the paradoxical signaling network structure (Figure \ref{fig:figure2}A) and multiple tiers within this network structure (Figure \ref{fig:figure2}B) allowed for temporal control of peak activities.

Upon \calcium influx, CaMKII is activated and further autophosphorylated resulting in a `molecular switch' that renders CaMKII active even after the \calcium pulse has dissipated \cite{Pi2008}.  Autophosphorylation of CaMKII-coupled with the Calcineurin-I1-PP1 mediated dephosphorylation events give rise to a paradoxical signaling network structure that can exhibit bistability (Figure \ref{fig:figure3}A). The bistability of CaMKII activation is shown in Figure \ref{fig:figures2}; the steady state CaMKII activity depends on the concentration of PP1. For low concentrations of PP1, CaMKII activity is sustained whereas for high concentration of PP1, CaMKII activity is transient (also see \cite{Pi2008}). Given the importance of CaMKII in synaptic function, it is not surprising that it can exhibit transient or long-term activation as shown in Figure \ref{fig:figure3}(B). In the spine, we find that GEF and GAP regulation by CaMKII results in a paradoxical network structure that controls the temporal activation of these components (Figure \ref{fig:figure4}A, B). 

Actin-mediated growth is a very important aspect of spine volume regulation \cite{Fischer1998,Matus2000}. Although the different aspects of actin regulation are coming to light, till date, there is no comprehensive model of how actin remodeling can affect the dendritic spine volume. We used a previously published model of actin remodeling in cell motility \cite{Tania2013} and applied it to spine dynamics (Figure \ref{fig:figure5}). We further linked the Rho activity to myosin activity and assumed that spine volume increase was proportional to the generation of new actin barbed ends and spine volume decrease is proportional to the myosin activity. The net result is spine volume increase and decrease that mimics the experimentally observed behavior \cite{Murakoshi2012,Kim2014,Murakoshi2011}. 

The usefulness of a modeling study is measured not just by its ability to explain experimentally observed behavior but also to generate experimentally testable predictions. Sensitivity analysis reveals the role of different components in the system (Table \ref{table:initial_index} and Figure \ref{fig:initial_conditions}). We outline below the set of experimentally testable predictions  from our model, which can be tested using a combination of silencing RNAs, gene knockouts, and pharmacological inhibitors. We also identify the current model limitations and scope for future work. 
%*\subsection*{Experimentally testable predictions}

 CaMKII dynamics are sensitive to many different components but primarily to the phosphastases. 
 \begin{itemize}
 	\item \textbf{CaMKII dynamics}: CaMKII dynamics are sensitive to many different components but primarily to the phosphastases. Phosphatases are often thought to be ubiquitous in their role in turning off phosphorylation mediated signaling activity. Here, we find that phosphatases, acting as an ultrasensitive cascade, coupled with autophosphorylation of CaMKII and autodephosphorylation of PP1 to result in a bistable phase profile (also seen in \cite{Zhabotinsky2000}. Sensitivity analysis shows that the dynamics of CaMKII are sensitive to PP1 and Calcineurin.  The model predicts that the time scale of CaMKII transience is extremely sensitive to phosphatase concentration (Figure \ref{fig:figures2}) and that the volume change of the spine is sensitive to the concentrations of calmodulin (Figure \ref{fig:figures3}B) and calcineurin (Figure \ref{fig:figures3}C). PP1 is thought to promote forgetting and inhibition of PP1, which is associated with sustained CaMKII activation, is thought to prolong memory \cite{genoux2002}. While there are many intermediate steps between CaMKII dynamics and memory formation, it is possible that the interaction between CaMKII and PP1 is a small but important step in governing learning and memory formation. 
 	
 	\item \textbf{Cdc42 dynamics}: Sensitivity analysis shows that Cdc42 dynamics are affected primarily by the concentrations of GEFs. The model predicts that depletion of Cdc42 will alter the dynamics of spine volume (Figure \ref{fig:figures4}) only slightly. This is because the barbed end generation is in the `coherent' regime \cite{Lacayo2007}. However Kim \textit{et al.} \cite{Kim2014}, showed that loss of Cdc42 may lead to defects in synaptic plasticity.   Since our model only focuses on single spine dynamics in a well-mixed condition, the actual dynamics at the whole cell level may be quite different from what we observe in the model.  

 	\item \textbf{Rho and myosin modification}: Since Rho modulates the myosin, Rho knockout leads to lack of myosin activity, resulting in a situation where the spine volume increases but decreases slowly or not at all (Figure \ref{fig:figures5}). While the role of myosin IIb has been tested using blebbistatin \cite{ryu2006}, we predict that upstream regulators of myosin activity will also alter spine compaction rates.The role of ROCK has been explored in \cite{Rex2010,Rex2009}, where the experiments show that Rho, ROCK, and myosin IIb are required for stable expression of LTP, but the role of these components in the early change in spine volume has not yet been tested.

 	\item \textbf{Cofilin regulation}: In our model, cofilin plays an important role in spine dynamics through its roles in actin severing and depolymerization. Our model predicts that a cofilin knockout will result in impaired actin dynamics that will disrupt the balance of actin remodeling during spine volume change (Figures \ref{fig:nuc_sev}, \ref{fig:figures4}) \cite{hotulainen2010,Bosch2014}. The effect of actin-mediated pushing is through barbed end generation, and the loss of cofilin removes the ability to sever existing filaments and generate new barbed ends. Experiments inhibiting cofilin showed that cofilin is highly enriched in the spine within 20s after stimulation and inhibition of cofilin using shRNA leads to a smaller enlargement of the spine \cite{Bosch2014}.
	
 	\item \textbf{Actin modulators}: To model actin dynamics we have extended a model developed by Tania et al \cite{Tania2013} to include both G- and F-actin dynamics. The model allows us to study the impact of pharmacological inhibitors on actin such as latrunculin and cytochalasin. The model predicts that latrunculin (which binds to G-actin and limits polymerization) will lead to a smaller increase in spine volume (Figure \ref{fig:figures4}D). Indeed, experiments showed that Latrunculin A when applied within 30s-2 min disrupts F-actin increases that are usually observed in spine expansion \cite{Rex2009}. On the other hand, removing actin capping also affects spine dynamics, resulting in increased barbed end production and larger spine volume (Figure \ref{fig:figures6}A). Therefore, filament capping is an important step in governing spine dynamics. Similar to Arp2/3, the kinetic parameter for the nucleation of new barbed ends had a small effect since barbed ends are also generated by severing (Figure \ref{fig:figures6} B, D). 
 \end{itemize}

%\subsection*{Sloppiness versus robustness}
A problem unique to large biochemical networks is parameter estimation. Recently, a comprehensive analysis of parametric sensitivity applied to many different systems biology models showed that there is a universal sloppiness associated with individual parameters in any model \cite{gutenkunst2007}.  However, we find that the interface between biochemical signaling and actin remodeling can be a source of robustness. That is, while the signaling dynamics is critical for initiating the actin remodeling events associated with spine dynamics, the actin remodeling events operate in the coherent regime, thereby reducing sensitivities to  parameter changes in the signaling network (Figure \ref{fig:kinetic_parameters}, Table \ref{table:kinetic_index}). Previously, we showed that for cell spreading, integrin signaling is required to initiate the actin remodeling but the spreading velocity and cell shape were primarily controlled by the actin-membrane interaction through the elastic Brownian ratchet \cite{Rangamani2011}. Here, we propose a similar mechanism for spine dynamics, where the \calcium-CaMKII signaling is required to initiate the biochemical activity associated with actin remodeling but the large concentrations of actin ensure that there is a robust response of the spine dynamics.  We compare our findings with the work of Bosch \textit{et al.} \cite{Bosch2014}, which study makes a key point -- namely that the `synaptic tag' designation is defined by the `increased binding capacity of the actin cytoskeleton' rather than any single molecule. This idea is in resonance with the central message of our study established by the sensitivity analysis that F-actin barbed ends  ensure a robust response in the `coherent' regime due to their abundance.

 %\subsection*{Perspectives and outlook}
To summarize, we present a dynamical systems model of the events that affect the transient spine dynamics. We identified that a simple module of paradoxical signaling can be used to explain the dynamics of CaMKII, actin remodeling, and spine volume change. We found that  once the barbed end generation is in the coherent regime, spine dynamics is robustly controlled by actin remodeling. Thus, the interface between signaling and actin barbed end generation is a source of natural robustness despite model sensitivity to kinetic parameters.
 
While our model has been able to explain the dynamics of spine volume change in the 4-5 minute time scale, we observe that our model can be enhanced in future versions. Currently, our model represents well-mixed dynamics. Spatial regulation of the Rho GTPases, and other components including cofilin, and myosin IIb, in the postsynaptic and neighboring spines have been observed in many studies \cite{Murakoshi2011}. Development of a spatio-temporal model that accounts for diffusive transport along with shape change of the spine is an important direction in our future work. Additionally, more complex models will be needed to identify cross-talk between different signaling components, antagonistic activities displayed by the same component at different concentrations and in different spatial compartments, the link between the dynamics of the dendritic spine and the dendritic shaft.

\subsubsection*{Acknowledgments}
This work was  supported by the National Institutes of Health Grant R01GM104979 to G.O. and the UC Berkeley Chancellor's Postdoctoral Fellowship, and the Air Force Office of Scientific Research award number FA9550-15-1-0124 to P.R. The authors would also like to thank Ms. Jasmine Nirody for help with extracting data from the time courses of the experiment. The Virtual Cell is supported by NIH Grant Number P41 GM103313 from the National Institute for General Medical Sciences.

\newpage
\renewcommand{\baselinestretch}{1}
\section*{Tables}
\begin{center}
\begin{table*}[!!h]
\centering
\caption{Characteristics of different species in the signaling cascade}
\renewcommand{\arraystretch}{1}
\begin{tabulary} {\textwidth}{LLLLL}
\hline
\hline
Species & Signal  & Signal &Time to peak(s) &  RMSE for\\
& exposure $I$  &Duration $\tau$(s) &  & model-experiment \\
& & & &comparison\\ [0.5ex]
\hline
CaMKII & 66.67 & 44.1 & 13 & 0.11 \\
Rho & 159.48 & 117.2& 48 & 0.04\\
Cdc42 & 167 & 130.31 & 51 & 0.25\\
Spine volume &184 &150.62  & 102 & 0.04\\
\hline
\end{tabulary}
\label{table:characteristics}
\end{table*}
\end{center}

\begin{center}
\begin{table*}[!!h]
\centering
\caption{Sensitivity to initial conditions}
\renewcommand{\arraystretch}{1.5}
\begin{tabulary} {\textwidth}{LLLLLL}
\hline
Index & Component & CaMKII & Cdc42 & Rho & Radius\\[0.5ex]
\hline
1 &Cofilin & \checkmark & \text{\sffamily X} & \text{\sffamily X} & \text{\sffamily X}\\
2& Arp23 & \checkmark & \text{\sffamily X} & \text{\sffamily X} & \text{\sffamily X}\\	
3 &CaN & \checkmark & \text{\sffamily X} & \text{\sffamily X} & \text{\sffamily X}\\
4&CaMKII-Factin  &\checkmark & \text{\sffamily X} & \checkmark & \text{\sffamily X}\\	
5& CaMKII-Gactin & \checkmark & \text{\sffamily X} & \checkmark	 & \text{\sffamily X}\\
6&Cdc42GTP & \checkmark & \checkmark & \text{\sffamily X} & \text{\sffamily X}\\	
7&Cdc42 GEF & \text{\sffamily X} & \text{\sffamily X} & \text{\sffamily X} & \text{\sffamily X}\\
8&	I1 & \text{\sffamily X} & \text{\sffamily X} & \checkmark & \checkmark\\	
9& LIMK & \checkmark	& \text{\sffamily X} & \text{\sffamily X} & \text{\sffamily X}\\
10& MLC & \text{\sffamily X} & \text{\sffamily X} & \text{\sffamily X} &\text{\sffamily X}\\
11& 	Myoppase active & \text{\sffamily X} & \text{\sffamily X} & \text{\sffamily X} & \text{\sffamily X}\\
12&	Myoppase & \checkmark & \text{\sffamily X} & \text{\sffamily X} & \text{\sffamily X}\\	
13& Ng & \checkmark & \text{\sffamily X} & \text{\sffamily X}	 & \text{\sffamily X}\\ 
14& PP1 & \checkmark & \text{\sffamily X} & \checkmark & \text{\sffamily X}\\	
15&RhoGAP & \text{\sffamily X} & \text{\sffamily X} & \text{\sffamily X} & \text{\sffamily X}\\
16&	RhoGEF & \text{\sffamily X} & \text{\sffamily X} & \text{\sffamily X} & \text{\sffamily X}\\
17&	RhoGDP & \text{\sffamily X} & \text{\sffamily X} & \text{\sffamily X} & \text{\sffamily X}\\
18&	ROCK & \text{\sffamily X} & \text{\sffamily X} & \checkmark & \checkmark\\
19&	SSh1 & \checkmark & \text{\sffamily X} & \text{\sffamily X} & \checkmark\\	
20& WASP & \checkmark & \checkmark & \text{\sffamily X} & \text{\sffamily X}\\
\hline
\hline
\end{tabulary}
\label{table:initial_index}
\end{table*}
\end{center}

\begin{landscape}
\begin{table}[]
\small
\centering
\caption{Sensitivity to kinetic parameters}
\label{table:kinetic_index}
\begin{tabular}{@{}lllllllllllll@{}}
\toprule
Index & Name & CaMKII & Cdc42 & Rho & Radius &  & Index & Name & CaMKII & Cdc42 & Rho & Radius \\ \midrule
1 & $k_{age}$ & \text{\sffamily X} & \text{\sffamily X} & \text{\sffamily X} & \text{\sffamily X} &  & 41 & $k_f$ ROCK inactivation & \text{\sffamily X} & \text{\sffamily X} & \text{\sffamily X} & \text{\sffamily X} \\
2 & $V_0$ & \text{\sffamily X} & \text{\sffamily X} & \text{\sffamily X} & \checkmark &  & 42 & $K_m$ for CaMKII activation & \text{\sffamily X} & \text{\sffamily X} & \text{\sffamily X} & \text{\sffamily X} \\
3 & $\omega$ & \text{\sffamily X} & \text{\sffamily X} & \text{\sffamily X} & \text{\sffamily X} &  & 43 & $K_{m1}$ for CaMKII autoactivation & \text{\sffamily X} & \text{\sffamily X} & \text{\sffamily X} & \text{\sffamily X} \\
4 & $k_{cap}$ & \text{\sffamily X} & \text{\sffamily X} & \text{\sffamily X} & \checkmark &  & 44 & $K_{m1}$  for PP1 activation by I1 & \text{\sffamily X} & \text{\sffamily X} & \text{\sffamily X} & \text{\sffamily X} \\
5 & $k_{deg}$ & \text{\sffamily X} & \text{\sffamily X} & \text{\sffamily X} & \text{\sffamily X} &  & 45 & $K_{m2}$ for PP1 autoactivation & \text{\sffamily X} & \text{\sffamily X} & \text{\sffamily X} & \text{\sffamily X} \\
6 & $k_{nuc}$ & \text{\sffamily X} & \text{\sffamily X} & \text{\sffamily X} & \text{\sffamily X} &  & 46 & $K_m$ for CaN activation by CaM & \text{\sffamily X} & \text{\sffamily X} & \text{\sffamily X} & \text{\sffamily X} \\
7 & $k_{sev}$ & \text{\sffamily X} & \text{\sffamily X} & \text{\sffamily X} & \text{\sffamily X} &  & 47 & $K_m$ for CaN inactivation & \text{\sffamily X} & \text{\sffamily X} & \text{\sffamily X} & \text{\sffamily X} \\
8 & $k_{shrink}$ & \text{\sffamily X} & \text{\sffamily X} & \text{\sffamily X} & \text{\sffamily X} &  & 48 & $K_m$ for CaMKII inactivation by CaM & \text{\sffamily X} & \text{\sffamily X} & \text{\sffamily X} & \text{\sffamily X} \\
9 & $\kappa$ & \text{\sffamily X} & \text{\sffamily X} & \text{\sffamily X} & \text{\sffamily X} &  & 49 & $K_m$ for CaMKII inactivation & \checkmark & \text{\sffamily X} & \checkmark & \text{\sffamily X} \\
10 & $k_f$ for Arp2/3 inactivation & \text{\sffamily X} & \text{\sffamily X} & \text{\sffamily X} & \text{\sffamily X} &  & 50 & $K_m$ Cdc42 activation & \text{\sffamily X} & \text{\sffamily X} & \text{\sffamily X} & \text{\sffamily X} \\
11 & $k$ for CaN activation by CaM & \text{\sffamily X} & \text{\sffamily X} & \text{\sffamily X} & \text{\sffamily X} &  & 51 & $K_m$ Cdc42GEFactivation & \text{\sffamily X} & \text{\sffamily X} & \text{\sffamily X} & \text{\sffamily X} \\
12 & $k_{cat}$ for PP1 activation by I1 & \text{\sffamily X} & \text{\sffamily X} & \text{\sffamily X} & \text{\sffamily X} &  & 52 & $K_m$ Cdc42GEF inactivation & \checkmark & \text{\sffamily X} & \text{\sffamily X} & \text{\sffamily X} \\
13 & $k_{cat}$ for PP1 autodephosphorylation & \text{\sffamily X} & \text{\sffamily X} & \text{\sffamily X} & \text{\sffamily X} &  & 53 & $K_m$ for Cdc42 hydrolysis & \text{\sffamily X} & \text{\sffamily X} & \text{\sffamily X} & \text{\sffamily X} \\
14 & $k_{cat}$ for CaMKII activation by CaM & \text{\sffamily X} & \text{\sffamily X} & \text{\sffamily X} & \text{\sffamily X} &  & 54 & $K_m$ cofilin activation SSH1 & \text{\sffamily X} & \text{\sffamily X} & \text{\sffamily X} & \text{\sffamily X} \\
15 & $k_{cat}$ CaMKII autophoshorylation & \checkmark & \text{\sffamily X} & \text{\sffamily X} & \text{\sffamily X} &  & 55 & $K_m$ I1 activation by CaN & \text{\sffamily X} & \text{\sffamily X} & \text{\sffamily X} & \text{\sffamily X} \\
16 & $k_{cat}$ CaMKII inactivation by PP1 & \checkmark & \text{\sffamily X} & \checkmark & \text{\sffamily X} &  & 56 & $K_m$ I1 inactivation & \text{\sffamily X} & \text{\sffamily X} & \text{\sffamily X} & \text{\sffamily X} \\
17 & $k_{cat}$ Cdc42 activation by GEF & \text{\sffamily X} & \text{\sffamily X} & \text{\sffamily X} & \text{\sffamily X} &  & 57 & $K_m$ LIMK activation by ROCK & \text{\sffamily X} & \text{\sffamily X} & \text{\sffamily X} & \text{\sffamily X} \\
18 & $k_{cat}$ Cdc42GEF activation by CaMKII & \text{\sffamily X} & \text{\sffamily X} & \text{\sffamily X} & \text{\sffamily X} &  & 58 & $k_m$ LIMK inactivation by SSH1 & \text{\sffamily X} & \text{\sffamily X} & \text{\sffamily X} & \text{\sffamily X} \\
19 & $k_{cat}$ Cdc42GEF inactivation by PP1& \text{\sffamily X} & \text{\sffamily X} & \text{\sffamily X} & \text{\sffamily X} &  & 59 & $K_m$ MLC dephosphorylation & \text{\sffamily X} & \text{\sffamily X} & \text{\sffamily X} & \text{\sffamily X} \\
20 & $k_{cat}$ Cdc42 hydrolysis & \text{\sffamily X} & \text{\sffamily X} & \text{\sffamily X} & \text{\sffamily X} &  & 60 & $K_m$ Myoppase activation & \text{\sffamily X} & \text{\sffamily X} & \text{\sffamily X} & \text{\sffamily X} \\
21 & $k_{cat}$ Cofilin activation by SSH1 & \text{\sffamily X} & \text{\sffamily X} & \text{\sffamily X} & \text{\sffamily X} &  & 61 & $K_m$ Myoppase inactivation & \text{\sffamily X} & \text{\sffamily X} & \text{\sffamily X} & \text{\sffamily X} \\
22 & $k_{cat}$ Cofilin inactivation by LIMK & \text{\sffamily X} & \text{\sffamily X} & \text{\sffamily X} & \text{\sffamily X} &  & 62 & $K_m$ PP1 inactivation by CaMKII & \text{\sffamily X} & \text{\sffamily X} & \text{\sffamily X} & \text{\sffamily X} \\
23 & $k_{cat}$ I1 inactivation by CaMKII & \text{\sffamily X} & \text{\sffamily X} & \text{\sffamily X} & \text{\sffamily X} &  & 63 & $K_m$ RhoGEF inactivation & \text{\sffamily X} & \text{\sffamily X} & \text{\sffamily X} & \text{\sffamily X} \\
24 & $k_{cat}$ LIMK activation by ROCK & \text{\sffamily X} & \text{\sffamily X} & \text{\sffamily X} & \text{\sffamily X} &  & 64 & $K_m$ RhoGAP activation by CaMKII & \text{\sffamily X} & \text{\sffamily X} & \text{\sffamily X} & \text{\sffamily X} \\
25 & $k_{cat}$ MLC phosphorylation by ROCK & \text{\sffamily X} & \text{\sffamily X} & \text{\sffamily X} & \text{\sffamily X} &  & 65 & $K_m$ RhoGAP inactivation by PP1& \text{\sffamily X} & \text{\sffamily X} & \text{\sffamily X} & \text{\sffamily X} \\
26 & $k_{cat}$ Myoppase dephosphorylation & \text{\sffamily X} & \text{\sffamily X} & \text{\sffamily X} & \text{\sffamily X} &  & 66 & $K_m$ RhoGEF activation by CaMKII& \text{\sffamily X} & \text{\sffamily X} & \text{\sffamily X} & \text{\sffamily X} \\
27 & $k_{cat}$ Myoppase inactivation by ROCK & \text{\sffamily X} & \text{\sffamily X} & \text{\sffamily X} & \text{\sffamily X} &  & 67 & $K_m$ RhoGTP by RhoGEF & \text{\sffamily X} & \text{\sffamily X} & \text{\sffamily X} & \text{\sffamily X} \\
28 & $k_{cat}$ PP1 inactivation by CaMKII & \text{\sffamily X} & \text{\sffamily X} & \text{\sffamily X} & \text{\sffamily X} &  & 68 & $K_m$ RhoGTP hydrolysis & \text{\sffamily X} & \text{\sffamily X} & \text{\sffamily X} & \text{\sffamily X} \\
29 & $k_{cat}$ RhoGAP activation by CaMKII & \text{\sffamily X} & \text{\sffamily X} & \text{\sffamily X} & \text{\sffamily X} &  & 69 & $K_m$ SSH1 dephosphorylation & \text{\sffamily X} & \text{\sffamily X} & \text{\sffamily X} & \text{\sffamily X} \\
30 & $k_{cat}$ RhoGEF activation by CaMKII & \text{\sffamily X} & \text{\sffamily X} & \text{\sffamily X} & \text{\sffamily X} &  & 70 & $K_m$ SSH1 phosphorylation by CaMKII & \text{\sffamily X} & \text{\sffamily X} & \text{\sffamily X} & \text{\sffamily X} \\
31 & $k_{cat}$ RhoGEF activity& \text{\sffamily X} & \text{\sffamily X} & \text{\sffamily X} & \text{\sffamily X} &  & 71 & $k_r$ for \calcium CaM binding & \text{\sffamily X} & \text{\sffamily X} & \text{\sffamily X} & \text{\sffamily X} \\
32 & $k_f$ \calcium CaM binding & \checkmark & \text{\sffamily X} & \text{\sffamily X} & \text{\sffamily X} &  & 72 & $k_r$ Arp2/3 activation & \text{\sffamily X} & \text{\sffamily X} & \text{\sffamily X} & \text{\sffamily X} \\
33 & $k_f$ Arp2/3 activation & \text{\sffamily X} & \text{\sffamily X} & \text{\sffamily X} & \text{\sffamily X} &  & 73 & $k_r$ CaM binding Ng & \text{\sffamily X} & \text{\sffamily X} & \text{\sffamily X} & \text{\sffamily X} \\
34 & $k_f$ CaM Ng binding& \text{\sffamily X} & \text{\sffamily X} & \text{\sffamily X} & \text{\sffamily X} &  & 74 & $k_r$ CaMKII binding F-actin & \text{\sffamily X} & \text{\sffamily X} & \text{\sffamily X} & \text{\sffamily X} \\
35 & $k_f$ CaMKII-F-actin binding & \text{\sffamily X} & \text{\sffamily X} & \text{\sffamily X} & \text{\sffamily X} &  & 75 & $k_r$ CaMKII G-actin & \text{\sffamily X} & \text{\sffamily X} & \text{\sffamily X} & \text{\sffamily X} \\
36 & $k_f$ CaMKII-G-actin binding & \text{\sffamily X} & \text{\sffamily X} & \text{\sffamily X} & \text{\sffamily X} &  & 76 & $k_r$ Cdc42 WASP bidning & \text{\sffamily X} & \text{\sffamily X} & \text{\sffamily X} & \text{\sffamily X} \\
37 & $k_f$ WASP activation by Cdc42 & \text{\sffamily X} & \text{\sffamily X} & \text{\sffamily X} & \text{\sffamily X} &  & 77 & $k_r$ MLCp basal activity & \text{\sffamily X} & \text{\sffamily X} & \text{\sffamily X} & \text{\sffamily X} \\
38 & $k_f$ MLC phosphorylation & \text{\sffamily X} & \text{\sffamily X} & \text{\sffamily X} & \text{\sffamily X} &  & 78 & $k_r$ Myoppase basal activity& \text{\sffamily X} & \text{\sffamily X} & \text{\sffamily X} & \text{\sffamily X} \\
39 & $k_f$ Myoppase basal activity & \text{\sffamily X} & \text{\sffamily X} & \text{\sffamily X} & \text{\sffamily X} &  & 79 & $k_r$ ROCK activation by Rho & \text{\sffamily X} & \text{\sffamily X} & \text{\sffamily X} & \text{\sffamily X} \\
40 & $k_f$ ROCK activation by Rho & \text{\sffamily X} & \text{\sffamily X} & \text{\sffamily X} & \text{\sffamily X} &  & 80 & $k_r$ ROCK inactivation & \text{\sffamily X} & \text{\sffamily X} & \text{\sffamily X} & \text{\sffamily X} \\ \bottomrule
\end{tabular}
\end{table}
\end{landscape}

\clearpage
%\section*{Figures}

\begin{figure*}
\centerline{\includegraphics[width=\textwidth]{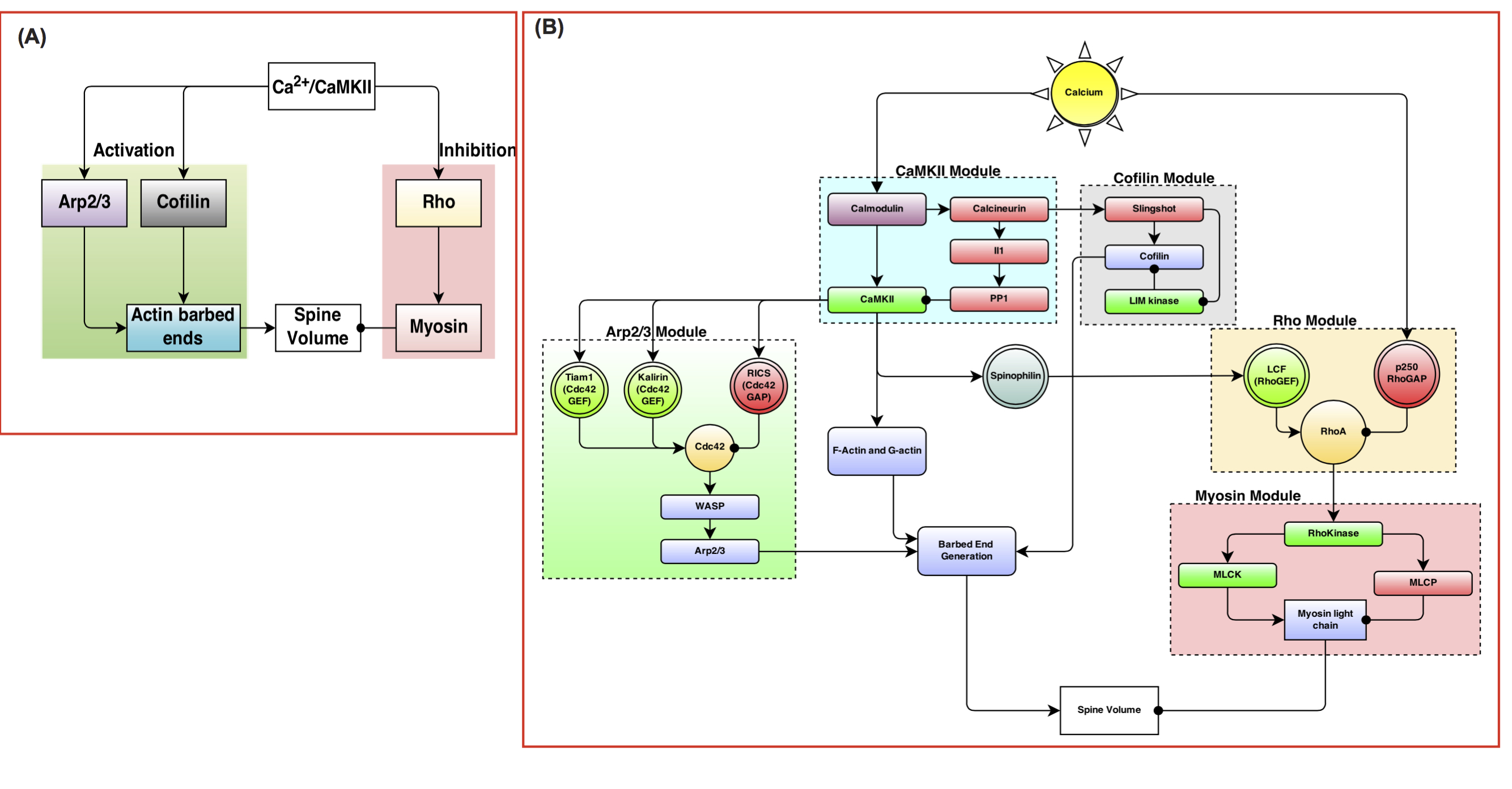}}
\caption{ (A) The events regulating transient spine enlargement can be thought of as modules that regulate key components. (B) The detailed biochemical network shows the interaction between different biochemical species. The network is constructed in modules, where each module contains one key component recognized as a key regulator of dendritic spine volume dynamics. The detailed table of reactions and parameters are provided in the tables in the supplementary material.}
\label{fig:figure1}
\end{figure*}

\begin{figure*}
\centerline{\includegraphics[width=\textwidth]{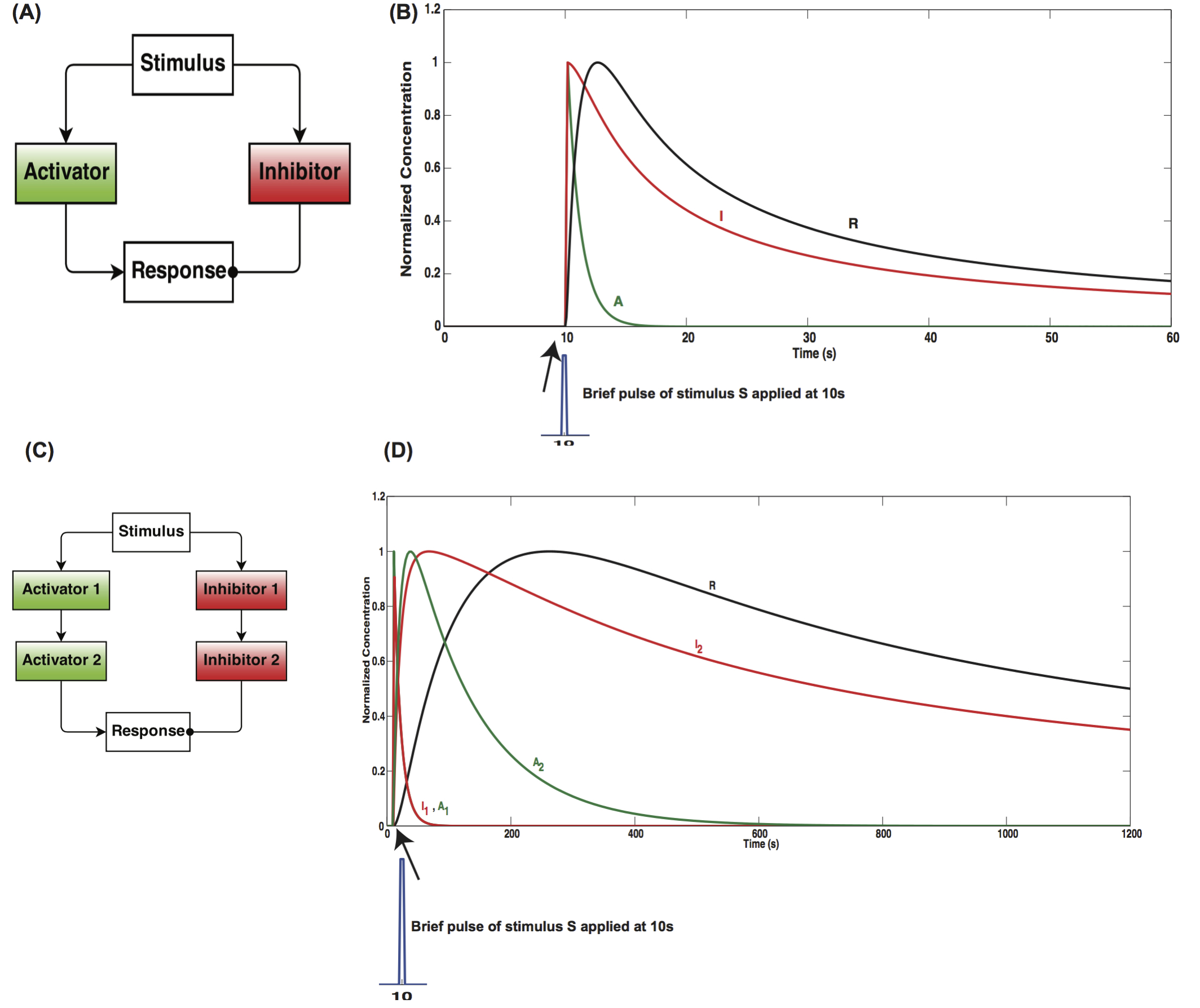}}
\caption{(A) An activation inhibition loop governed by the same stimulus S leads to the dynamics of the response R as shown in (B). This is an example of paradoxical signaling. In this network structure, the temporal dynamics of the response $R$ can be regulated by a biexponential function (Figure \ref{fig:figures1}) in the supplemental material, where the exponential associated with the increase of $R$ is controlled by the activator and the exponential associated with the decay of $R$ is controlled by the inhibitor. Tuning the effect of the exponentials is enough to move the peak of the response curve (see supplemental material for a detailed discussion).  The multi-level activation inhibition loop in Figure \ref{fig:figure1}B can be represented by a toy model as shown in (C). The time course of the response R is dependent on the number of upstream tiers, essentially resulting in biexponential functions of exponentials. The first level of activator and inhibitor show an early response, the second activator and inhibitor show an intermediate time course and the response $R$ shows a delay compared to the stimulus presented at 10 s. This indicates that multi-tiered activation-inhibition networks of paradoxical signaling are sufficient to control the temporal evolution of the response.}
\label{fig:figure2}
\end{figure*}

\begin{figure*}
\centerline{\includegraphics[scale=0.2]{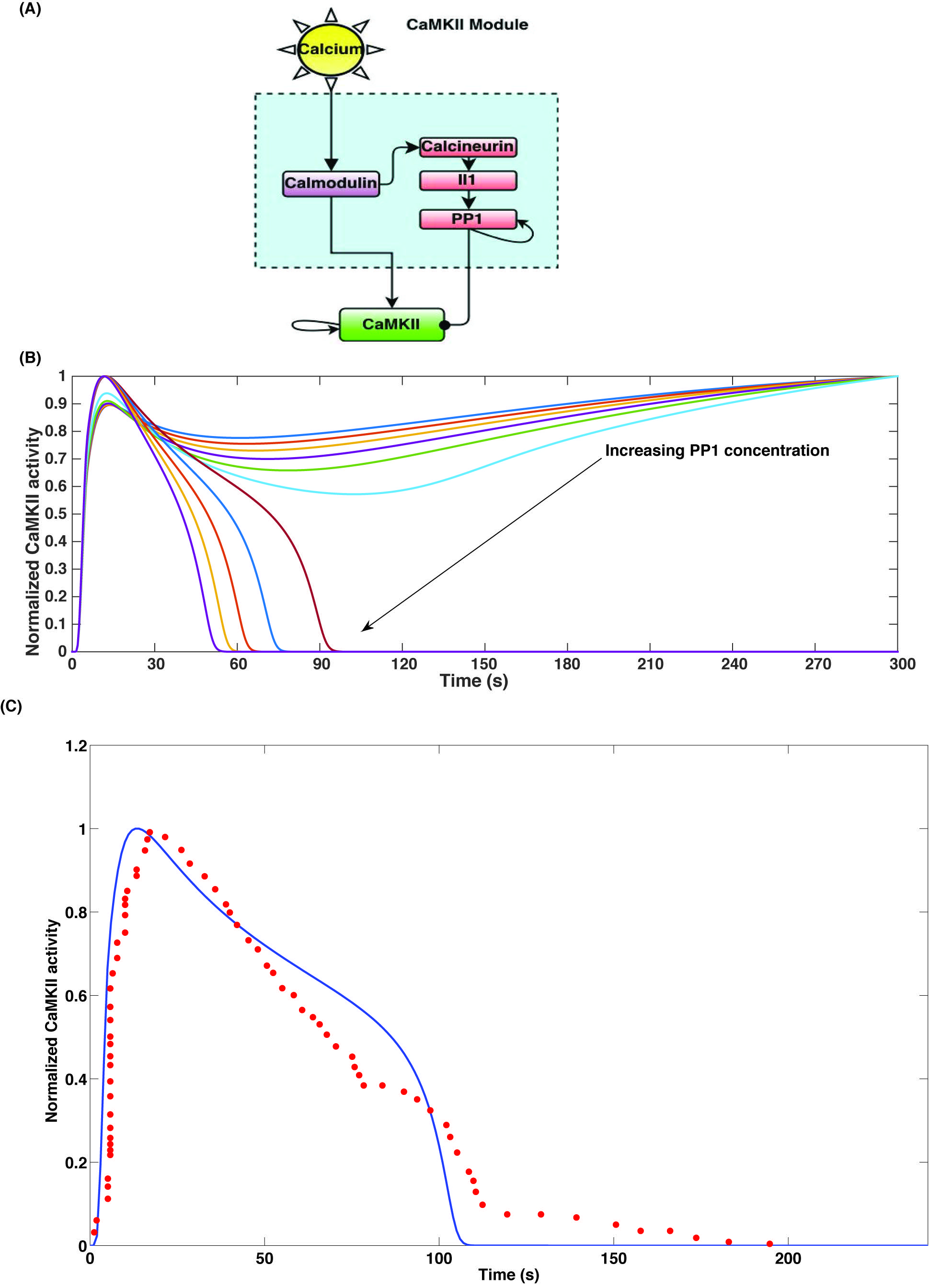}}
\caption{ (A) CaMKII module: CaMKII is activated by \calcium-Calmodulin. Calmodulin also activates an ultrasensitive phosphatase cascade, which includes Calcineurin, Il1, and PP1. As a result, the temporal dynamics of CaMKII activation is tightly regulated by both \calcium-Calmodulin led CaMKII autophosphorylation and PP1-mediated dephosphorylation. (B) The coupling between the phosphatase cascade and the autophosphorylation events results in a bistable response of CaMKII. The different curves in the plot are CaMKII dynamics in response to increasing PP1 concentration. PP1 concentration was varied from $0$ to $1$ $\mu M$. Increasing PP1 concentration results in switching from sustained CaMKII activation to transient CaMKII activation.  (C) In our model, PP1 initial concentration of $0.36 \mu M$ results in CaMKII dynamics that matches closely with experimental data. The blue solid line is the simulation data and the red filled circles are the experimental data. The experimental data was extracted from Figure 1D of Murakoshi \textit{et al.} \cite{Murakoshi2012} using the digitize package in `R'. The root mean square error was 0.11 for n=63 data points from experiments.}
\label{fig:figure3}
\end{figure*}

\begin{figure*}
\centerline{\includegraphics[scale=0.2]{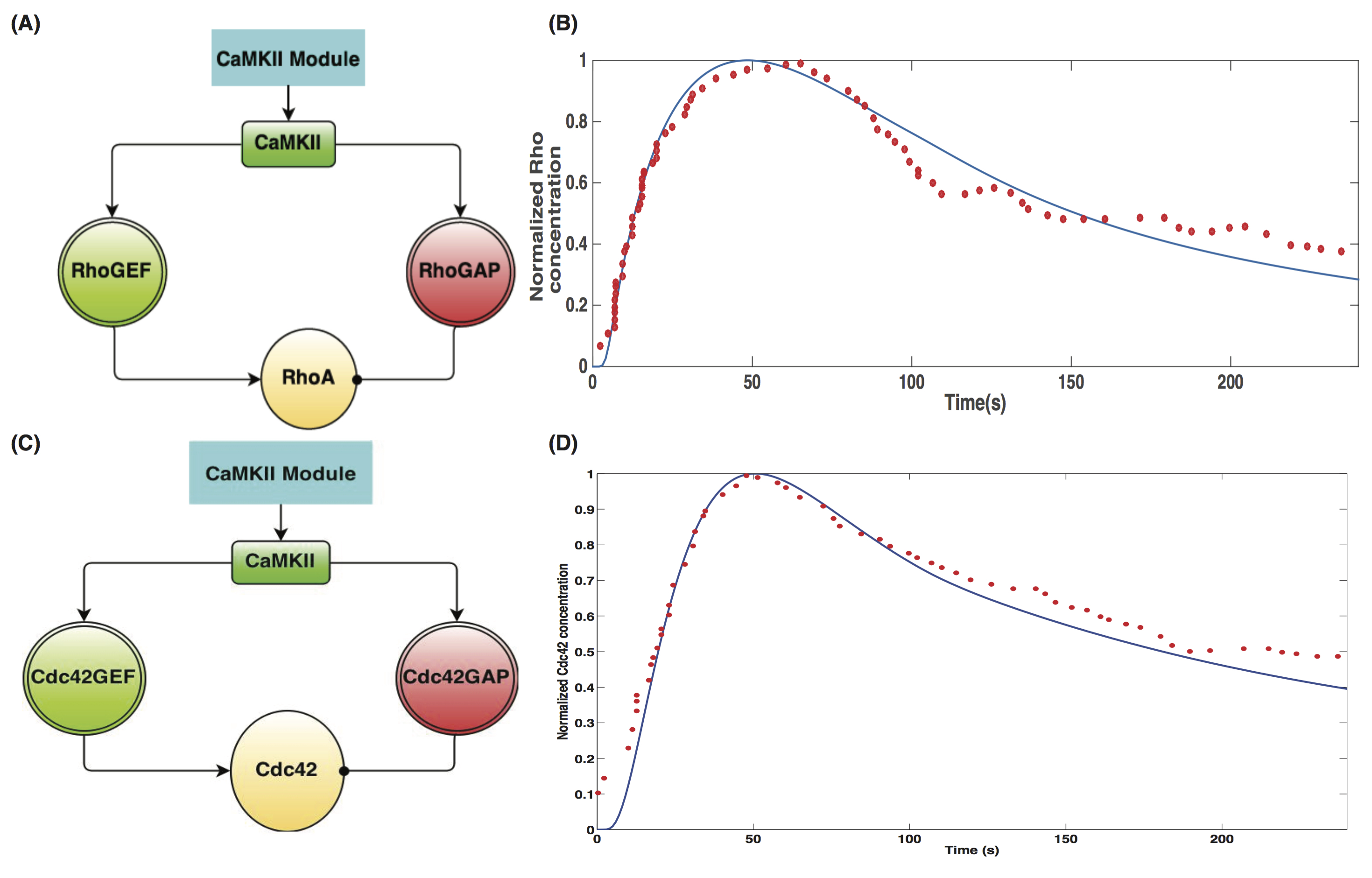}}
\caption{ CaMKII activates the small Rho-GTPases, RhoA and Cdc42, by regulating the activation of the GEFs and the GAPs. The resulting network structure within these modules is similar to the activation-inhibition module of paradoxical signaling shown in Figure \ref{fig:figure2}A. (A) RhoGTPase activation Module. (B) Comparison of experimental data from Figure3D of Murakoshi \textit{et al.} \cite{Murakoshi2012} with model simulations for RhoGTP. The blue solid line is the simulation data and the red filled circles are the experimental data.  The root mean square error was 0.04 for n=65 data points. (C) Cdc42 activation module, (D) Comparison of experimental and simulation data for Cdc42. The blue solid line is the simulation data and the red filled circles are the experimental data. The experimental data was obtained from Figure 1D of \cite{Murakoshi2012}. The root mean square error was $0.25$ for n=58 data points from experiments.}
\label{fig:figure4}
\end{figure*}

\begin{figure*}
\centerline{\includegraphics[scale=0.19]{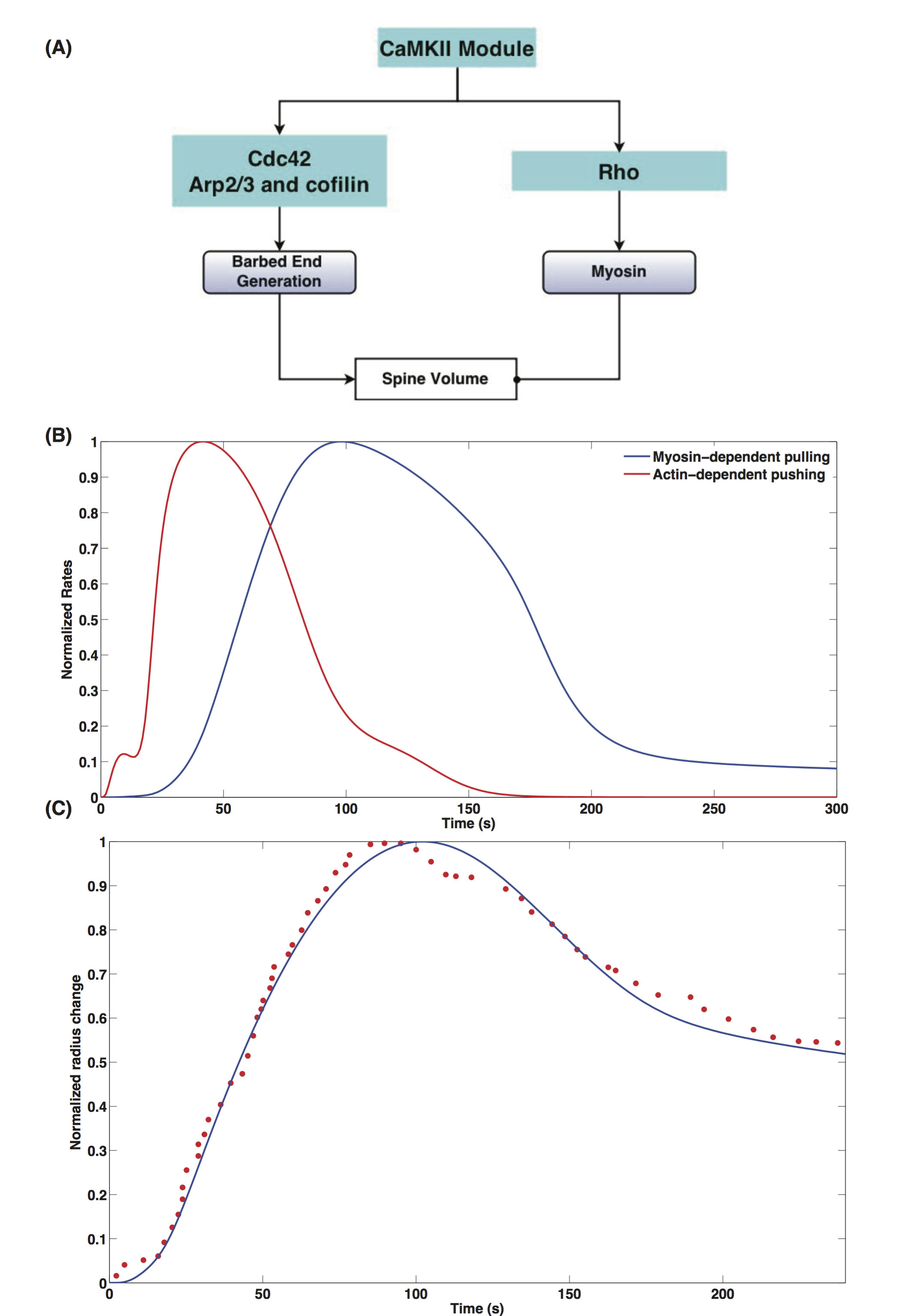}}
\caption{Spine size regulation by \calcium-CaMKII can be represented as a multi-tier activation-inhibition loop of paradoxical signaling as shown in Figure \ref{fig:figure2}(C). (A) Spine size regulation module showing the interaction between the upstream signaling modules CaMKII, Rho, Cdc42 and the barbed end generation and myosin activation. (B) Normalized rates of actin-dependent spine growth due to pushing of barbed ends and myosin dependent spine volume decrease, (C) Comparison of experimental and simulation data for normalized spine volume. The blue solid line is the simulation data and the red filled circles are the experimental data. The experimental data was obtained from Figure 1D of \cite{Murakoshi2012}. The root mean squared error is 0.046, with n=62 data points. }
\label{fig:figure5}
\end{figure*}

\begin{figure*}
\centerline{\includegraphics[scale=0.2]{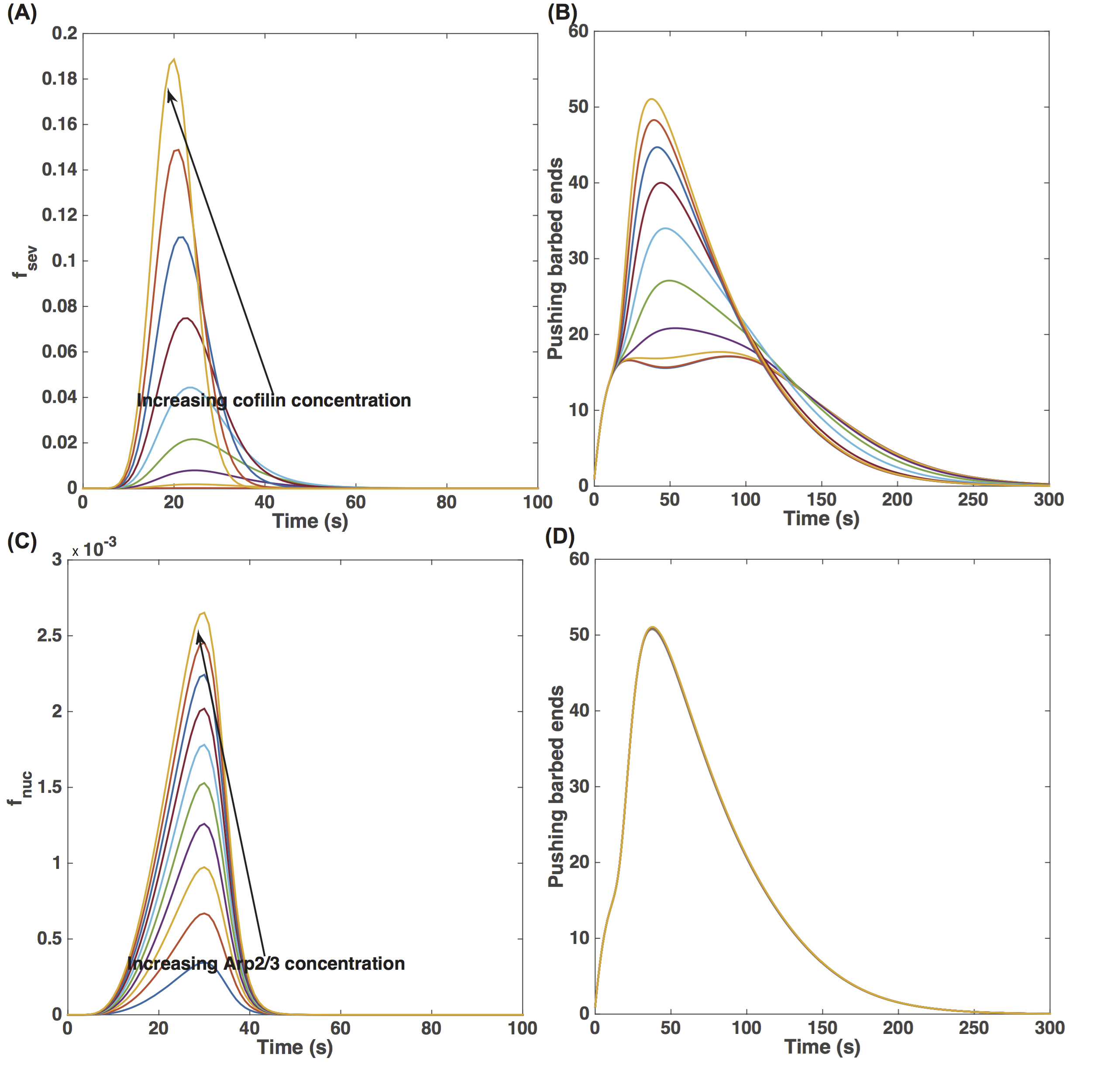}}
\caption{Effect of cofilin and Arp2/3 concentrations on pushing barbed end generation. (A) Increasing cofilin concentration leads to a cooperative increase in severing rate, $f_{sev}$, and (B) subsequently an increase in pushing barbed end production. (C) Increasing Arp2/3 concentration on the other hand does not lead to a cooperative increase in nucleation rate, $f_{nuc}$, and (D) the impact of Arp2/3 concentration on pushing barbed end concentration is negligible.}
\label{fig:nuc_sev}
\end{figure*}

\begin{figure*}
\centerline{\includegraphics[scale=0.45]{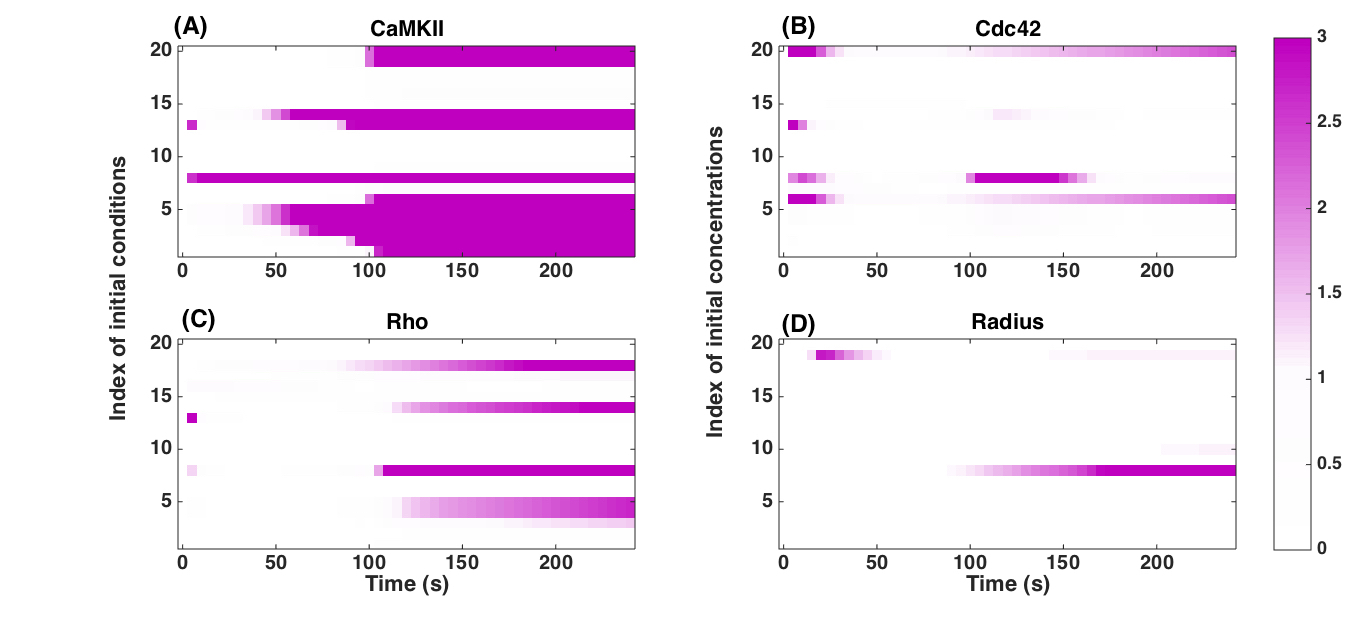}}
\caption{Sensitivity to initial conditions. We calculated the local sensitivity coefficient (Eq. \ref{eq:sensitivity}) with respect to the initial concentrations of the different components in the model for (A) CaMKII concentration, (B) Cdc42-GTP concentration, (C) Rho-GTP concentration, and (D) Spine radius as a function of time. White in the plots indicates that the sensitivity is zero. Any colors towards the purple end of the color range can be interpreted as high sensitivity. The color maps show the absolute scale of $S_{i,j}$. The index of concentrations is given in Table \ref{table:initial_index}.}
\label{fig:initial_conditions}
\end{figure*}

\begin{figure*}
\centerline{\includegraphics[scale=0.45]{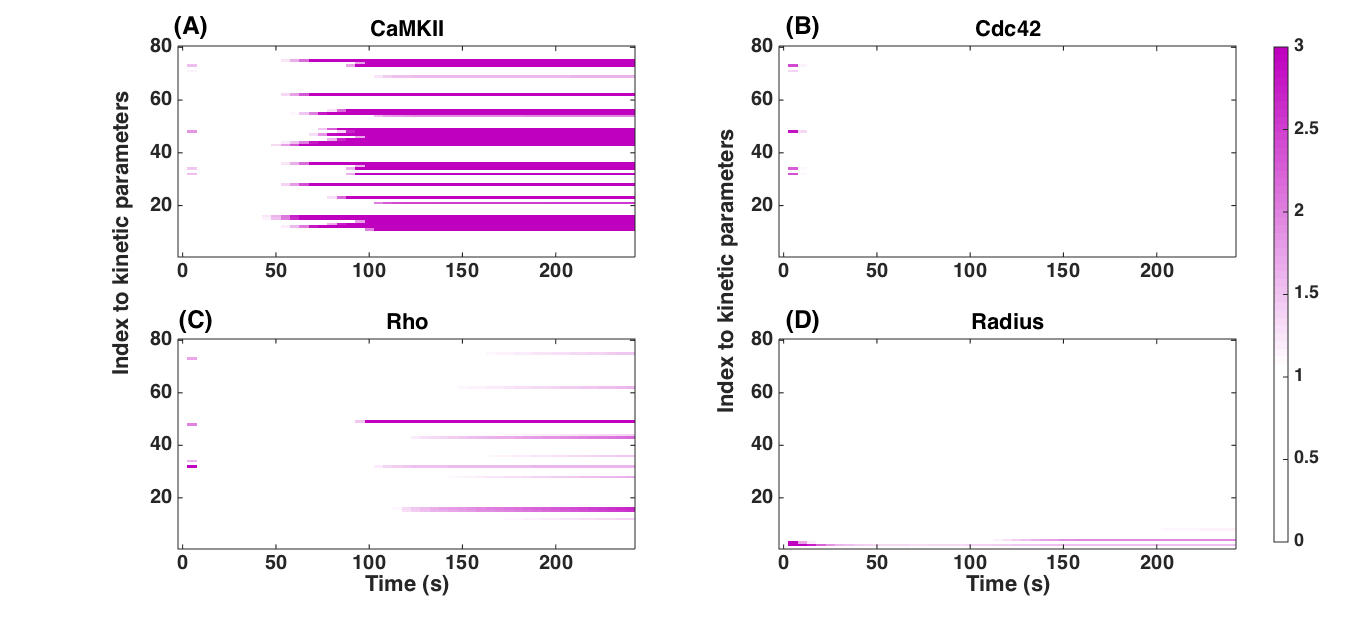}}
\caption{Sensitivity to kinetic parameters. We calculated the local sensitivity coefficient (Eq. \ref{eq:sensitivity}) with respect to the kinetic parameters in the model for (A) CaMKII concentration, (B) Cdc42-GTP concentration, (C) Rho-GTP concentration, and (D) Spine radius as a function of time. Note that the model is robust to changes in many kinetic parameters. White in the plots indicates that the sensitivity is zero. Any colors towards the purple end of the color range can be interpreted as high sensitivity. The color maps show the absolute scale of $S_{i,j}$.  The index of kinetic parameters is given in Table \ref{table:kinetic_index}.}
\label{fig:kinetic_parameters}
\end{figure*}

\clearpage
%\bibliographystyle{unsrt}
%\bibliography{ref}

% Bibliography
%\bibliography{pnas-sample}

\newpage

\setcounter{equation}{0}
\setcounter{figure}{0}
\setcounter{table}{0}
\setcounter{page}{1}
\setcounter{table}{0}
\renewcommand{\thetable}{S\arabic{table}}
\renewcommand{\theequation}{S\arabic{equation}}
\renewcommand{\thefigure}{S\arabic{figure}}

\begin{center}
\LARGE Supplemental Material for \\
Paradoxical signaling regulates structural plasticity in dendritic spines
\\~\\
\large Padmini Rangamani$^{1*}$,  Michael G. Levy$^2$,  Shahid M. Khan$^3$,  and George Oster$^4$
\\~\\
$^1$ Department of Mechanical and Aerospace Engineering,\\
 University of California San Diego, La Jolla CA 92093\\
$^2$ Biophysics Graduate Program, University of California Berkeley, \\
Berkeley CA 94720\\
$^3$ Molecular Biology Consortium, Lawrence Berkeley National Laboratory, \\
Berkeley CA 94720\\
$^4$ Department of Molecular and Cell Biology, University of California Berkeley, \\
Berkeley CA 94720.\\
$^*$ To whom correspondence should be addressed: padmini.rangamani@eng.ucsd.edu.

\end{center}

\newpage

\section*{A brief introduction to modeling chemical reactions}
\subsection*{Mass-action kinetics}
  We generate an ordinary differential equation (ODE) for each species using mass-action kinetics for each reaction. Under mass action, the rate of a chemical reaction is  proportional to the product of the reactant concentrations raised to the power of their stoichiometric coefficients. For example, consider the one-reaction system:
\begin{equation}
X + Y \rightleftharpoons  2Z,
\end{equation} 
where the forward and backward rates are $k_1$ and $k_2$. The differential equations describing the dynamics of species $X,Y$, and $Z$ under mass-action kinetics are:
\begin{align}
\frac{d[X]}{dt} &= k_2 [Z]^2 - k_1[X][Y]\\
\nonumber
\frac{d[Y]}{dt} &= k_2 [Z]^2 - k_1[X][Y]\\
\nonumber
\frac{d[Z]}{dt} &= k_1 [X][Y] - k_2[Z]^2.
\end{align}
\subsection*{Michaelis-Menten kinetics}
When the reaction is catalyzed by an enzyme, with kinetic properties $k_{cat}$ and $K_M$, 
\begin{equation}
\nonumber
S \xrightarrow[\quad]{E} P	%\ch{S ->[E] P},
\end{equation} 
then the reaction rate is given by
\begin{equation}
	\frac{d[S]}{dt} = -\frac{k_{cat}[E][S]}{K_M+[S]}=\frac{d[P]}{dt}
\end{equation}
\section*{Paradoxical signaling}

\subsection*{One-Tier Model of Paradoxical Signaling}
The phenomenological model underlying the activation-inhibition loop shown in Figure \ref{fig:figure2}A can be written as a system of ordinary differential equations (ODEs) to track the temporal dynamics of the response function $R$ due to time-dependent stimulus input $S(t)$ (details in Table \ref{table:onetier}). The activator $A$ and inhibitor $I$, transmit information from the stimulus to the response. $^*$ indicates the activated fraction of the species.  Using mass-action kinetics, we can formulate a system of ODEs to represent the change in concentration of each species over time.  Since $A(t)$ and $I(t)$ are immediate effectors of the stimulus $S(t)$, they demonstrate a rapid increase in activation and decay exponentially (Figure \ref{fig:figure2}A, B). The response $R(t)$ shows a slightly delayed peak compared to $A(t)$ and $I(t)$, which is consistent with degree of separation of $R(t)$ from $S(t)$. The dynamics of each of these components follows a biexponential function, which is sufficient to explain the time scales of activation and inhibition. These results suggest that changing the parameters of the above model is sufficient to reproduce all the results of the temporal dynamics of spine volume change coupled with CaMKII and RhoGTPases \cite{Murakoshi2012}. 

Under the assumption of mass action kinetics, the rate of a chemical reaction is  proportional to the product of the reactant concentrations raised to the power of their stoichiometric coefficients. For the simple paradoxical signaling network shown in Figure \ref{fig:figure2}A, the dynamics can be modeled using the reactions in Table \ref{table:onetier}. $k_i$'s represent the forward reaction rates. For simplicity, we ignore the backward reaction rate here; including the reverse reaction will not affect the qualitative behavior of the system. $S$ is a controlled, time-dependent input, a pulse function, to model the burst of calcium into the spine. The system of ordinary differential equations resulting from this system are given in Eq. \ref{eqn:AI} and were solved using the ode45 routine in MATLAB (Mathworks, Natick, MA) (Figure \ref{fig:figures1}B). It is straightforward to note that this simple system gives rise to the behaviors observed in the spine dynamics but also in many other signaling systems \cite{Alon2007,Hart2012,Hart2013}. 

\begin{center}
\begin{table*}[!!h]
\centering
\caption{Reactions for a one-tier activation-inhibition loop}
\renewcommand{\arraystretch}{1.5}
\begin{tabular} {l l l l }
\hline
\multicolumn{4}{l}{Initial concentrations (all in $\mu$M): [A] =$A_0$; [I] = $I_0$; [R] = $R_0$}\\
\hline
Reaction &  k$_{\text{on}}$ & Notes\\ [0.5ex]
\hline
A + S $\rightarrow$ A* & $k_{1}$  & Stimulus activates activator\\
A* + R $\rightarrow$ R* & $k_{3}$   & Response is activated by A\\
I + S $ \rightarrow$ I* & $k_{2}$  & Stimulus activates inhibitor\\
R* + I* $\rightarrow$ R & $k_{4}$  & Response in inhibited by I\\
\hline
\end{tabular}
\label{table:onetier}
\end{table*}
\end{center}

The corresponding differential equations are given as

\begin{eqnarray}
\frac{d[A]}{dt}=-k_1[A]S(t)\\
\nonumber
\frac{d[A^*]}{dt}=k_1[A]S(t)-k_2[A^*][R]\\
\nonumber
\frac{d[I]}{dt}=-k_3[I]S(t)\\
\nonumber
\frac{d[I^*]}{dt}=k_3[I]S(t)-k_4[I^*][R^*]\\
\nonumber
\frac{d[R]}{dt}=-k_2[A^*][R]+k_4[I^*][R^*]\\
\nonumber
\frac{d[R^*]}{dt}=k_2[A^*][R]-k_4[I^*][R^*]
\label{eqn:AI}
\end{eqnarray}

%\begin{eqnarray}

%\end{eqnarray}

\subsection*{Multi-tier model of paradoxical signaling}

\begin{center}
\begin{table*}[!!h]
\centering
\caption{Reactions for a multi-tier activation-inhibition loop}
\renewcommand{\arraystretch}{1.5}
\begin{tabular} {l l l l }
\hline
\multicolumn{4}{l}{Initial concentrations (all in $\mu$M): [A] =$A_0$; [I] = $I_0$; [R] = $R_0$}\\
\hline
Reaction &  k$_{\text{on}}$ & Notes\\ [0.5ex]
\hline
A$_1$ + S $\rightarrow$ A$_1$* & $k_{1}$  & Stimulus activates activator 1\\
A$_1$* + A2 $\rightarrow$ A$_2$* & $k_{2}$  & Activator 1 activates activator 2\\
A$_2$* + R $\rightarrow$ R* & $k_{3}$   & Response is activated by activator 2\\
I$_1$ + S $ \rightarrow$ I$_1$* & $k_{4}$  & Stimulus activates inhibitor 1 \\
I$_2$ + I$_1$* $ \rightarrow$ I$_2$* & $k_{5}$  & Inhibitor 1 activates inhibitor 2\\
R* + I$_2$* $\rightarrow$ R & $k_{6}$  & Response in inhibited by inhibitor 2\\
\hline
\end{tabular}
\label{table:multitier}
\end{table*}
\end{center}
The reactions for the multi-tier paradoxical signaling are given in Table \ref{table:multitier}, where $k_i$'s represent the forward reaction rates. $S$ is a controlled, time-dependent input, a pulse function, to model the burst of calcium into the spine. The system of ordinary differential equations resulting from this system are given in Eq. \ref{eqn:AI1} and were solved using the ode45 routine in MATLAB (Mathworks, Natick, MA). 

\begin{eqnarray}
\frac{d[A_1]}{dt}=-k_1[A_1]S(t)\\
\nonumber
\frac{d[A_1^*]}{dt}=k_1[A_1]S(t)-k_2[A_1^*][A_2]\\
\nonumber
\frac{d[A_2^*]}{dt}=k_2[A_1^*][A_2]-k_3[A_2^*][R]\\
\nonumber
\frac{d[I_1]}{dt}=-k_4[I_1]S(t)\\
\nonumber
\frac{d[I_1^*]}{dt}=k_4[I_1]S(t)-k_5[I_1^*][I2]\\
\nonumber
\frac{d[I_2^*]}{dt}=k_5[I_1^*][I2]-k_6[I_2^*][R^*]\\
\nonumber
\frac{d[R]}{dt}=-k_3[A_1^*][R]+k_6[I_2^*][R^*]\\
\nonumber
\frac{d[R^*]}{dt}=k_2[A^*][R]-k_4[I^*][R^*]
\label{eqn:AI1}
\end{eqnarray}

We summarize our observations from the toy models here: in the case of a simple biexponential function Eq.\ref{eq:biexponential}, it is straightforward to see how the two exponents, $a$ and $b$, characterize the inhibition and activation dynamics respectively. In the case of dynamics regulated by large signaling networks, $a$ and $b$ are no longer constants but regulated by upstream processes and activity of proteins. Nonetheless, this simple function gives us large insight into how homeostasis might come about. Further extending this idea to that of a one-tier paradoxical signaling model and subsequently the two tier model gives us a way of understanding how the kinetic parameters influence the time courses (Figure \ref{fig:figures1}B). Effectively, the large network shown in Figure \ref{fig:figure1}C is a larger scale system with the same dynamic behavior. 

\section*{Model development for spine volume change}

\subsection*{CaMKII Module}
The reactions in the CaMKII module are based on the events outlined in the Introduction section of the main text. NMDA receptor activation leads to a \calcium pulse, which then binds to Calmodulin, resulting in a calcium-calmodulin complex. This complex is key for the activation of CaMKII by binding to the kinase and relieving the autoinhibition. The structural details of how this happens are discussed in \cite{rosenberg2005,goldberg1996} and are beyond the scope of the current work. An additional key step in the activation of CaMKII is autophosphorylation. To represent these kinetics, we have used the model developed by Pi and Lisman \cite{Pi2008}. Calcium-calmodulin also activates calcineurin, also known as PP3. Calcineurin acts through a cascade of phosphatases I1 and PP1 to dephosphorylate CaMKII \cite{mulkey1994}. The calmodulin levels are regulated by a protein called neurogranin, which is found in large quantities in the brain \cite{zhabotinsky2006}. The kinetics of PP1 activity are described in \cite{Pi2008}. 

\subsection*{Cdc42 Module}
Cdc42 is a small RhoGTPase, which plays an important role in governing the actin remodeling events. It is required for the activation of WASP through PIP2 and the subsequent downstream activation of Arp2/3 \cite{}. The activity of Cdc42 is regulated by Guanine nucleotide exchange factors (GEFs) and GTPase activating proteins (GAPs). There are a large number of GEFs and GAPs present in the brain and their activity is known to be controlled directly or indirectly though CaMKII. We assumed that the GEFs and GAPs were activated by CaMKII (based on \cite{okamoto2009}) and were inactivated by PP1. Since phosphatases are known to be promiscuous \cite{virshup2009}, this assumption is justified. The GTP bound Cdc42 then activates WASP and Arp2/3 \cite{Rangamani2011,marchand2001,higgs2000}. 

\subsection*{Cofilin Module}
Cofilin is an important regulator of actin dynamics; by severing actin filaments, it plays an important role in recycling the actin \cite{sarmiere2004}. Cofilin is known to be negatively regulated by phosphorylation mediated by LIM kinase \cite{endo2003} and activated by dephosphorylation by SSH1 \cite{endo2003}. We use the module developed in \cite{Rangamani2011} to model cofilin dynamics. SSH1 itself is activated by calcineurin \cite{wang2005} and inhibited by CaMKII. LIM kinase is activated by Rho kinase, ROCK \cite{amano2010}. The cofilin that is activated by these events leads to the severing of actin filaments and is modeled using $f_{sev}$ (Table \ref{table:Actin}) based on the model presented in \cite{Tania2013}.

\subsection*{Actin module and pushing velocity}
We use the well-mixed model developed by Tania \textit{et. al.} \cite{Tania2013} with a few modifications, to model the dynamics of the barbed end generation from Arp2/3 and cofilin activity. The phosphorylation of CaMKII releases F-actin and G-actin in the spine, which are then free to generate barbed ends through filament nucleation and severing. The degradation and aging of filaments results in regeneration of G-actin. The functional forms of severing and nucleation are taken from \cite{Tania2013}. The barbed ends are capped at a fixed rate $k_{cap}$. Not all barbed ends generate a pushing velocity; the relationship between the pushing barbed ends $Bp$, and the total barbed ends is derived from the conservation conditions in \cite{Tania2013}. As explained in \cite{Tania2013}, Eq. \ref{eq:vmb} ensures that spine expansion is initiated only when barbed ends have built up sufficiently. The actin-mediated spine growth events are modeled as an elastic Brownian ratchet \cite{mogilner2003,mogilner1996}. The difference between our model and the one published in \cite{Tania2013} is that the species involved in the actin module are all regulated by upstream signaling regulated by CaMKII.

The relationship between pushing barbed ends, $Bp$, and the membrane velocity has been derived in \cite{Lacayo2007,Tania2013}, and is used here to calculate the radius of the growing spine in response to the actin remodeling events. This relationship is given as
\begin{equation}
V_{mb}=V_0\frac{Bp}{Bp+\phi \exp(\omega/Bp)}
\label{eq:vmb}
\end{equation}
Here, we used the values of $\phi=10/\mu m$, and $\omega=50$ per $\mu m$. \cite{Tania2013}. $\phi$ is the geometric parameter used in computing membrane protrusion rate and $\omega$ is the physical parameter describing the membrane resistance \cite{Tania2013,Lacayo2007}. This density-velocity relationship has a biphasic behavior -- for a small number of barbed ends, the membrane resistance limits the velocity, explained as `decoherent' regime in \cite{Lacayo2007} and for large barbed end density, or the `coherent' regime, the protrusion rate is not sensitive to the number of barbed ends. 

\subsection*{Rho-Myosin Module}
Rho is known to regulate ROCK and therefore myosin activity. In cell motility models, Rho is known to play an important role in contractility. Here, we apply the idea of myosin-mediated contraction that is governed by RhoGTP activity. Rho is a small GTPase and as in the Cdc42 module, we assume that the GEF and GAP activity are regulated by the CaMKII module. RhoGTP then activates ROCK, which is known to activate both myosin light chain kinase and myosin light chain phosphatase \cite{kaneko2012}. The reactions and reaction rates for ROCK activation of myosin light chain are based on  the model presented in \cite{kaneko2012}.

\subsection*{Spine radius}
We propose the following equation for the dynamics of the spine volume.
\begin{equation}
\frac{dR}{dt}=V_{mb}-k_{shrink}[MLC^*]R
\end{equation}
The spine growth velocity is assumed to depend on the pushing velocity generated by actin and the shrinkage of the spine is proportional to the amount of myosin light chain that is phosphorylated and the radius of the spine. A similar model force dependence on myosin concentration for contractility was proposed in \cite{barnhart2015}. 

\subsection*{Actin depolymerization or myosin-mediated contraction?}
 Actin depolymerization is another key factor in actin remodeling. An open question in the field is whether increasing actin depolymerization is sufficient to induce spine contraction.  In order to study the effect of actin depolymerization, we included another term in the conversion of F-actin to G-actin, which had the rate $k_{depol}[F-actin]$. The number of pushing barbed ends is affected by depolymerization (Figure \ref{fig:figures7}A), however for rates of depolymerization as observed in experiments \cite{Pollard1986} and used in other models \cite{Mogilner1996}, the spine dynamics is not greatly affected (Figure \ref{fig:figures7}B). More importantly, our conclusion that if the barbed ends are sufficient to drive the velocity to a coherent regime, then the spine dynamics is robust, still holds. Also, if the depolymerization rate is very high, there are not enough pushing barbed ends to generate a protrusive force (Figure \ref{fig:figures8}). This means that increasing actin depolymerization can decrease the growth velocity ($V_{mb}$ of the spine but not exert a contractile force. 
Therefore, when the net actin polymerization rate is reduced, as long as cofilin can generate barbed ends in the `coherent' regime and myosin can generate contractile forces, then we can capture the observed dynamics. There are many factors that contribute to barbed end generation -- polymerization, nucleation, and severing. If polymerization rate decreases, then the number of barbed ends with decrease, slowing down the velocity to zero and the spine volume will plateau out.  But to shrink, an active inward pulling force is needed. We demonstrate that here with a simple system.
\begin{equation}
\frac{dR_{spine}}{dt}=V_{spine}-V_{\text{pulling by myosin}}
\end{equation}

Suppose that there is no pulling by myosin and the decrease in polymerization results in $V_{spine}=0$. Then, we are left with $\frac{dR}{dt}=0$ whose solution is a constant radius. Therefore, for the radius to shrink, an active pulling force is needed (Figure \ref{fig:figures9}B).

\subsection*{Kinetic parameters}

One of the challenges of developing models of signaling networks is choice of kinetic parameters. As the system of reactions gets more complex, it is harder to identify with certainty what a certain kinetic parameter value would be. One way to address this issue is to match the dynamics from simulations with experimental measurements of input-output relationships \cite{neves2011}. In this case, the data that is available is for the four key readouts in the spine and is available as percent change or normalized value. This makes it harder for us to identify kinetic parameters. We overcame this challenge by selecting kinetic parameters for the various reactions from the literature where possible and indicate it in the Tables. In some cases, the values of reaction rate constants were not available and these parameters were fit to match the experimental time course. 
The units of the kinetic parameters are as follows: all first order reaction rates and $k_{cat}$s have units of $s^{-1}$, second order reaction rates have units of $\mu M s^{-1}$, and $K_m$ has units of $\mu M$. Barbed ends $B$ and pushing barbed ends $B_p$ have units of $\#/\mu m$ and velocities $V_0$ and $V_{mb}$ have units of $\mu m s^{-1}$. $l$ is the scale factor for converting units of actin concentration and has units of $\mu M\cdot\mu m$, and $\kappa$ is the scale factor for converting concentrations to units of barbed ends and has units of $\#/\mu m^2\cdot \mu M$ \cite{Tania2013}. 
\section*{Extracting experimental data from published work}
The experimental data was extracted from Figure 3D of \cite{Murakoshi2012} using the digitize package in the software `R' \cite{poisot2011}.

\section*{Model access in VCell}
The simulations of the full network shown in Figure \ref{fig:figure1}C were carried in the Virtual Cell program (http://www.nrcam.uchc.edu). The model is named `Spine Model Final' and is available under the publicly shared models with the username `prangamani'. Complete instructions on how access publicly shared models can be found at the Virtual Cell homepage. A teaching resource on how to develop models in Virtual Cell is given in \cite{neves2011a}.
\newpage
\section*{Tables}
The references in these tables indicate either the source of the reaction and/or the source for the parameters. Often interaction parameters are given in terms of dissociation constants and these values were used to obtain the forward and backward rates for mass-action kinetics. In other cases, parameters were obtained from other models or assumed such that they would fit the experimental data. 
\begin{center}
\begin{table*}[!!h]
\centering
\caption{Reactions for the activation of CaMKII}
\renewcommand{\arraystretch}{1.5}
\begin{tabulary} {\textwidth}{LLLL}
\hline
\hline
Reaction & Reaction flux & Kinetic Parameters & Reference\\ [0.5ex]
\hline
3\calcium + CaM $\rightarrow$ Ca$\cdot$CaM &   $k_f$[\calcium]$^3$[CaM]-$k_r$[Ca$\cdot$ CaM] & $k_f=7.75$, $k_r=1$ &  \cite{Pi2008} \\
Ng + CaM $\rightarrow$ CaM$\cdot$Ng & $k_f$[Ng][CaM]-$k_r$[Ca$\cdot$ CaM]& $k_f=5$, $k_r=1$ & \cite{zhabotinsky2006}\\
CaMKII + F-actin $\rightarrow$CaMKII$\cdot$F-actin &$k_f$[CaMKII][F-actin]-$k_r$[CaMKII]$\cdot$[F-actin] &$k_f=1$, $k_r=4$ & \cite{sanabria2009}\\
CaMKII + G-actin $\rightarrow$CaMKII$\cdot$G-actin &$k_f$[CaMKII][G-actin]-$k_r$[CaMKII]$\cdot$[G-actin] &$k_f=1$, $k_r=4$ & \cite{sanabria2009}\\ 
CaMKII $\rightarrow$ CaMKIIp & $\frac{k_{cat1}\text{[Ca$\cdot$CaM]}^4\text{[CaMKII]}}{K_{m1}^4+\text{[Ca$\cdot$CaM]}^4}+\frac{k_{cat2}\text{[CaMKIIp]}\text{[CaMKII]}}{K_{m2}+\text{[CaMKII]}}$ & $k_{cat1}=120$, $K_{m1}$=4, \\
& & $k_{cat2}=1$, $K_{m2}=10$ &\cite{Pi2008}\\
CaMKIIp $\rightarrow$ CaMKII & $\frac{k_{cat}\text{[PP1*]}\text{[CaMKIIp]}}{K_m+\text{[CaMKIIp]}}$ & $k_{cat}=15$, $K_m=3$ & \cite{Pi2008} \\
CaN $\rightarrow$ CaN* & $\frac{k\text{[Ca$\cdot$CaM]}^4\text{[CaN]}}{K_m^4+\text{[Ca$\cdot$CaM]}^4}$ & $k_{cat}=127$, $K_m$=0.34 & \cite{zhabotinsky2006} \\
CaN* $\rightarrow$ CaN& $\frac{k_{cat}\text{[CaMKIIp]}\text{[CaN]}}{K_m+\text{[CaN]}}$ & $k_{cat}=0.34$, $K_m=127$ & \cite{zhabotinsky2006} \\
I1 $\rightarrow$ I1* & $\frac{k_{cat}\text{[CaN*]}\text{[I1]}}{K_m+\text{[I1]}}$ & $k_{cat}=0.034$, $K_m$=4.97 & \cite{zhabotinsky2006} \\
I1* $\rightarrow$ I1 & $\frac{k_{cat}\text{[CaMKIIp]}\text{[I1*]}}{K_m+\text{[I1*]}}$ & $k_{cat}=0.0688$, $K_m=127$ & \cite{zhabotinsky2006} \\
PP1 $\rightarrow$ PP1* & $\frac{k_{cat1}\text{[I1*]}\text{[PP1]}}{K_{m1}+\text{[PP1]}}+\frac{k_{cat2}\text{[PP1*]}\text{[PP1]}}{K_{m2}+\text{[PP1]}}$ & $k_{cat1}=50$, $K_{m1}$=80, \\
& & $k_{cat2}=2$, $K_{m2}=80$ &\cite{Pi2008}\\
PP1* $\rightarrow$ PP1 & $\frac{k_{cat}\text{[CaMKIIp]}\text{[PP1*]}}{K_m+\text{[PP1*]}}$ & $k_{cat}=0.07166$, $K_m=4.97$ & \cite{zhabotinsky2006} \\
\hline
\end{tabulary}
\label{table:CaMKII}
\end{table*}
\end{center}

\begin{center}
\begin{table*}[!!h]
\centering
\caption{Reactions for the activation of Cdc42 and Arp2/3}
\renewcommand{\arraystretch}{1.5}
\begin{tabulary} {\textwidth}{LLLL}
\hline
\hline
Reaction & Reaction flux & Kinetic Parameters & Reference\\ [0.5ex]
\hline
Cdc42-GEF $\rightarrow$ Cdc42-GEF* &   $\frac{k_{cat}\text{[CaMKIIp]}\text{[Cdc42-GEF]}}{K_m+\text{[Cdc42-GEF]}}$ & $k_{cat}=0.01$, $K_m=1.0$ & \cite{zhang1997,fleming1999}, parameters estimated to match experimental time course \\
Cdc42-GEF* $\rightarrow$ Cdc42-GEF &   $\frac{k_{cat}\text{[PP1*]}\text{[Cdc42-GEF*]}}{K_m+\text{[Cdc42-GEF*]}}$ & $k_{cat}=0.01$, $K_m=1.0$ & \cite{zhang1997,fleming1999}, parameters estimated to match experimental time course \\
Cdc42-GDP$\rightarrow$ Cdc42-GTP &   $\frac{k_{cat}\text{[Cdc42-GEF*]}\text{[Cdc42-GDP]}}{K_m+\text{[Cdc42-GDP]}}$ & $k_{cat}=0.75$, $K_m=1.0$ & \cite{zhang1997}, parameters estimated to match experimental time course  \\
Cdc42-GTP$\rightarrow$ Cdc42-GDP &   $\frac{k_{cat}\text{[GAP*]}\text{[Cdc42-GTP]}}{K_m+\text{[Cdc42-GTP]}}$ & $k_{cat}=0.1$, $K_m=1.0$ & \cite{zhang1997}, parameters estimated to match experimental time course \\
GAP$\rightarrow$ GAP* &   $\frac{k_{cat}\text{[CaMKIIp]}\text{[GAP]}}{K_m+\text{[GAP]}}$ & $k_{cat}=0.01$, $K_m=1.0$ & parameters estimated to match experimental time course \\
GAP*$\rightarrow$ GAP &   $\frac{k_{cat}\text{[PP1*]}\text{[GAP*]}}{K_m+\text{[GAP*]}}$ & $k_{cat}=0.01$, $K_m=1.0$ & parameters estimated to match experimental time course \\
Cdc42-GTP + WASP $\rightarrow$ WASP* &$k_{f}\text{[Cdc42-GTP][WASP]}-k_{r}\text{[WASP*]}$ & $k_{f}=0.02$, $k_r=0.001$ & \cite{marchand2001,higgs2000,Rangamani2011} \\
Arp2/3 + WASP* $\rightarrow$ Arp2/3* & $k_{f}\text{[Arp2/3][WASP*]}-k_{r}\text{[Arp2/3*]}$ & $k_{f}=0.1$, $k_r=0.0$ & \cite{marchand2001,higgs2000,Rangamani2011} \\
\hline
\end{tabulary}
\label{table:Cdc42}
\end{table*}
\end{center}

\begin{center}
\begin{table*}[!!h]
\centering
\caption{Reactions for the activation of Cofilin}
\renewcommand{\arraystretch}{1.5}
\begin{tabulary} {\textwidth}{LLLL}
\hline
\hline
Reaction & Reaction flux & Kinetic Parameters & Reference\\ [0.5ex]
\hline
SSH1 $\rightarrow$ SSH1* &   $\frac{k_{cat}\text{[CaN*]}\text{[SSH1]}}{K_m+\text{[SSH1]}}$ & $k_{cat}=0.34$, $K_m=4.97$ & \cite{zhabotinsky2006} \\
SSH1* $\rightarrow$ SSH1 &   $\frac{k_{cat}\text{[CaMKII*]}\text{[SSH1*]}}{K_m+\text{[SSH1*]}}$ & $k_{cat}=127$, $K_m=0.34$ & \cite{Pi2008} \\
LIMK$\rightarrow$ LIMK* &   $\frac{k_{cat}\text{[ROCK*]}\text{[LIMK]}}{K_m+\text{[LIMK]}}$ & $k_{cat}=0.9$, $K_m=0.3$ & \cite{trauger2002} \\
LIMK*$\rightarrow$ LIMK &   $\frac{k_{cat}\text{[SSH1*]}\text{[LIMK*]}}{K_m+\text{[LIMK*]}}$ & $k_{cat}=0.34$, $K_m=4.0$ & SSH1 was assumed to have similar kinetics as CaN; \cite{zhabotinsky2006} \\
Cofilin$\rightarrow$ Cofilin* &   $\frac{k_{cat}\text{[SSH1*]}\text{[Cofilin]}}{K_m+\text{[Cofilin]}}$ & $k_{cat}=0.34$, $K_m=4.0$ & SSH1 was assumed to have similar kinetics as CaN; \cite{zhabotinsky2006} \\
Cofilin*$\rightarrow$ Cofilin &   $\frac{k_{cat}\text{[LIMK*]}\text{[Cofilin*]}}{K_m+\text{[Cofilin*]}}$ & $k_{cat}=0.34$, $K_m=4.0$ & LIM Kinase was assumed to have similar parameters as CaMKII; \cite{Pi2008} \\
\hline
\end{tabulary}
\label{table:Cofilin}
\end{table*}
\end{center}

\begin{center}
\begin{table*}[!!h]
\centering
\caption{Reactions for the generation of actin barbed ends}
\renewcommand{\arraystretch}{1.5}
\begin{tabulary} {\textwidth}{LLLL}
\hline
\hline
Reaction & Reaction flux & Kinetic Parameters & Reference\\ [0.5ex]
\hline
F-new-actin $\rightarrow$ F-actin &   $k_{age}\text{[F-new]}$ & $k_{age}=0.001$ & \cite{Tania2013} \\
F-actin $\rightarrow$ G-actin &   $f_{sev}+k_{deg}\text{[F-actin]}+k_{age}\text{[F-actin]}$ & $k_{sev}=0.001$, $C_0=0.1$, $l=0.255$ & \cite{Tania2013} \\
& $f_{sev}=\frac{C_0k_{sev}\text{[Cofilin*]}^4l\text{[F-actin]}}{C_0^4}$& $k_{age}=0.1$, $k_{deg}=0.1$ & \\
F-actin+G-actin+Arp23*$\rightarrow$ F-actin &   $f_{nuc}=\frac{k_{nuc}\text{[Arp23*]}\text{[F-actin][G-actin]}l}{K_m+\text{[Arp23*]}}$ & $k_{nuc}=60$, $K_m=2$, $l=0.255$ & \cite{Tania2013} \\
$\rightarrow$ B &   $\kappa(f_{sev}+f_{nuc})-k_{cap}B$ & $k_{cap}=0.04$, $\kappa=106$ & \cite{Tania2013} \\
$\rightarrow$ Bp &   $(V_0-V_{mb})B-k_{cap}Bp$ & $V_0=0.1$, $k_{cap}=0.04$ & \cite{Tania2013} \\
\hline
\end{tabulary}
\label{table:Actin}
\end{table*}
\end{center}

\begin{center}
\begin{table*}[!!h]
\centering
\caption{Reactions for the activation of Rho and Myosin}
\renewcommand{\arraystretch}{1.5}
\begin{tabulary}  {\textwidth}{LLLL}
\hline
\hline
Reaction & Reaction flux & Kinetic Parameters & Reference\\ [0.5ex]
\hline
Rho-GEF $\rightarrow$ Rho-GEF* &   $\frac{k_{cat}\text{[CaMKIIp]}\text{[Rho-GEF]}}{K_m+\text{[Rho-GEF]}}$ & $k_{cat}=0.01$, $K_m=1.0$ & Est. \\
Rho-GEF* $\rightarrow$ Rho-GEF &   $\frac{k_{cat}\text{[PP1*]}\text{[Rho-GEF*]}}{K_m+\text{[Rho-GEF*]}}$ & $k_{cat}=0.1$, $K_m=1.0$ & Est. \\
Rho-GDP$\rightarrow$ Rho-GTP &   $\frac{k_{cat}\text{[Rho-GEF*]}\text{[Rho-GDP]}}{K_m+\text{[Rho-GDP]}}$ & $k_{cat}=0.75$, $K_m=1.0$ & Est. \\
Rho-GTP$\rightarrow$ Rho-GDP &   $\frac{k_{cat}\text{[GAP*]}\text{[Rho-GTP]}}{K_m+\text{[Rho-GTP]}}$ & $k_{cat}=0.1$, $K_m=1.0$ & Est. \\
Rho-GTP + ROCK $\rightarrow$ ROCK* &   $k_f[Rho-GTP][ROCK]-k_r[ROCK*]$ & $k_{f}=0.02$, $k_r=0.001$ & \cite{Rangamani2011} \\
MyoPpase$\rightarrow$ MyoPpase*&   $k_f[MyoPpase]+\frac{k_{cat}\text{[MyoPpase*]}\text{[MyoPpase]}}{K_{m}+\text{[MyoPpase]}}$ & $k_{f}=0.01$, $k_{cat}=3$, $K_{m}=16$ &\cite{kaneko2012}\\
MyoPpase* $\rightarrow$ MyoPpase & $\frac{k_{cat}\text{[ROCK*]}\text{[MyoPpase*]}}{K_m+\text{MyoPpase*]}}$ & $k_{cat}=2.357$, $K_m=0.1$ & \cite{kaneko2012} \\
MLC$\rightarrow$ MLC*&   $k_f[MLC]+\frac{k_{cat}\text{[ROCK*]}\text{[MLC]}}{K_{m}+\text{[MLC]}}$ & $k_{f}=0.01$, $k_{cat}=1.8$, $K_{m}=2.47$ &\cite{kaneko2012}\\
MLC* $\rightarrow$ MLC & $\frac{k_{cat}\text{[MyoPPase*]}\text{[MLC*]}}{K_m+\text{MLC*]}}$ & $k_{cat}=1$, $K_m=16$ & \cite{kaneko2012} \\

\hline
\end{tabulary}
\label{table:Rho}
\end{table*}
\end{center}

\begin{center}
\begin{table*}[!!h]
\centering
\caption{Initial conditions (only the species with non-zero initial conditions are shown)}
\renewcommand{\arraystretch}{1}
\begin{tabulary} {\textwidth}{LLLL}
\hline
\hline
Species & Initial concentration ($\mu M$) & Notes and References\\ [0.5ex]
\hline
CaMKII-F-actin & 10 & Assumed based on \cite{okamoto2009}; varied in our simulations (Figure \ref{fig:figures3}) \\
CaMKII-G-actin & 10 & Assumed based on \cite{okamoto2009}; varied in our simulations (Figure \ref{fig:figures3}) \\
Calcineurin & 1 & \cite{zhabotinsky2006}; varied in our simulations (Figure \ref{fig:figures3})\\
Calmodulin & 10 & \cite{zhabotinsky2006}; varied in our simulations (Figure \ref{fig:figures3})\\
Neurogranin & 20 & \cite{zhabotinsky2006}\\
I1 & 1.8 & \cite{zhabotinsky2006}; varied in our simulations (Figure \ref{fig:figures2})\\
LIM kinase & 2 & Assumed\\
Myoppase$^*$ &	 	0.1 & \cite{kaneko2012} \\
RhoGEF	 &	0.1 & Assumed\\
RhoGAP	 &	0.1 & Assumed\\
Cdc42GEF &	 	0.1 & Assumed \\
PP1	 &	0.27 &Estimated by parameter variation (Figure \ref{fig:figures2})\\
ROCK &	 	1.0 & Assumed; varied in our simulations (Figure \ref{fig:figures5})\\
RhoGDP &	 	1.0 & Assumed; varied in our simulations (Figure \ref{fig:figures5})\\
WASP	 &	1.0 & Assumed\\
Arp2/3	 &	1.0 & Assumed; varied in our simulations (Figure \ref{fig:figures4})\\
Cdc42GDP &	 	1.0 & Assumed; varied in our simulations (Figure \ref{fig:figures4})\\
Bp	 &	1.0 & Set to a small non-zero value to initiate simulations\\
Myoppase	 & 	1.1 & \cite{kaneko2012}\\
SSH1	 &	2.0 & Assumed\\
Cofilin	 &	2.0 & Assumed; varied in our simulations (Figure \ref{fig:figures4})\\
MLC	 &	5.0 &\cite{kaneko2012}; varied in our simulations (Figure \ref{fig:figures5})\\
B	 &	30.0 & \cite{Tania2013}; set to a small resting value to indicate that most of the actin is bundled to CaMKII\\
\hline
\end{tabulary}
\label{table:initial_conditions}
\end{table*}
\end{center}

\clearpage
\section*{Figures}

\begin{figure}[!!h]
\centerline{\includegraphics[width=\textwidth]{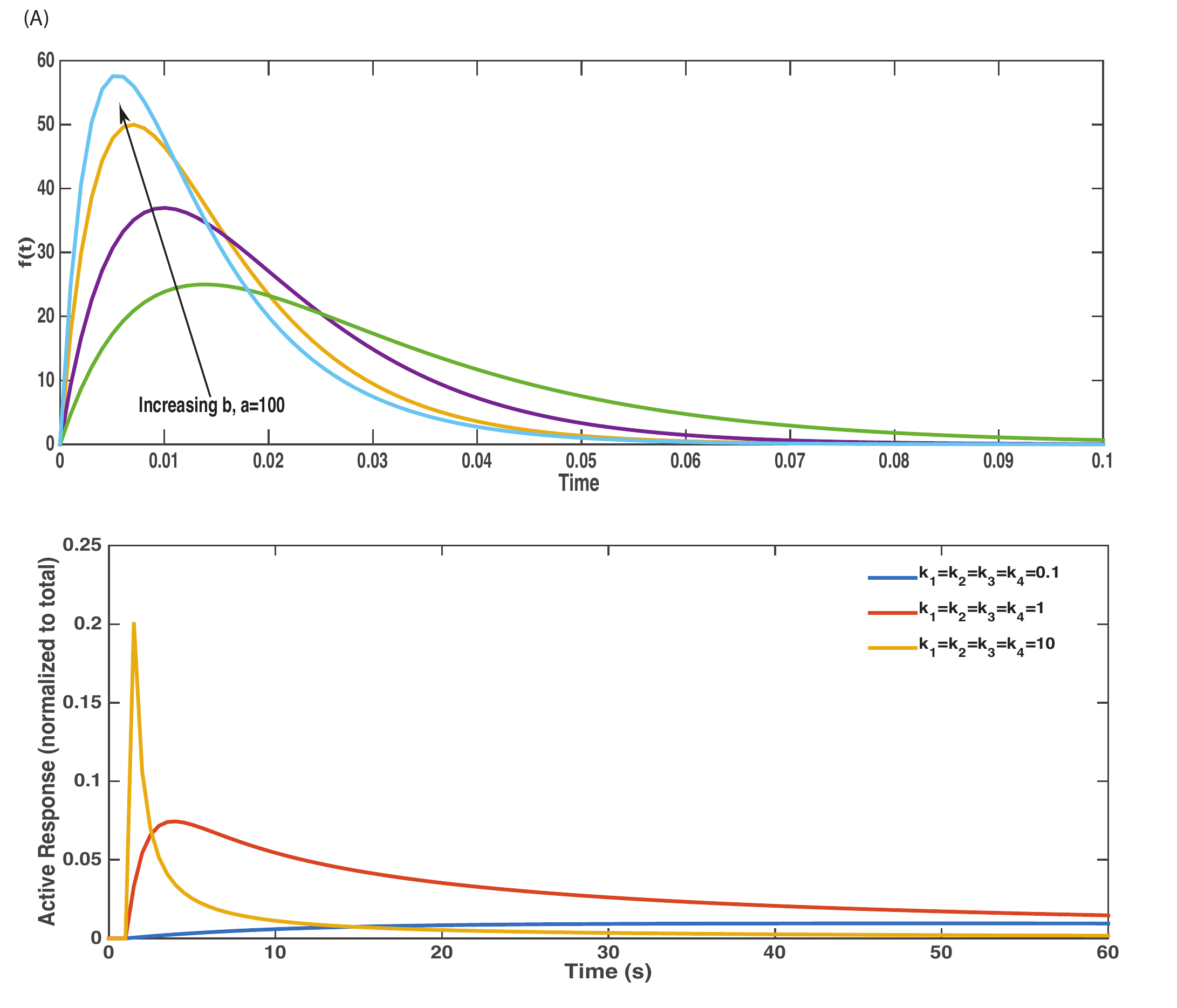}}
\caption{(A) The parameters of a biexponential function $f(t)$ can be tuned to change the dynamics of the function. This simple model shows how dynamics of the response function can be controlled by simple parameters. By tuning the parameters, $a$ and $b$, of the function, we can capture the fast response that is like that of CaMKII and a slow-response that is like the spine volume. (B) The role of parameters in governing the dynamics of the simple paradoxical signaling module is shown. Changing the parameters captures the full range of behavior.}
\label{fig:figures1}
\end{figure}

\begin{figure}[!!h]
\centerline{\includegraphics[width=\textwidth]{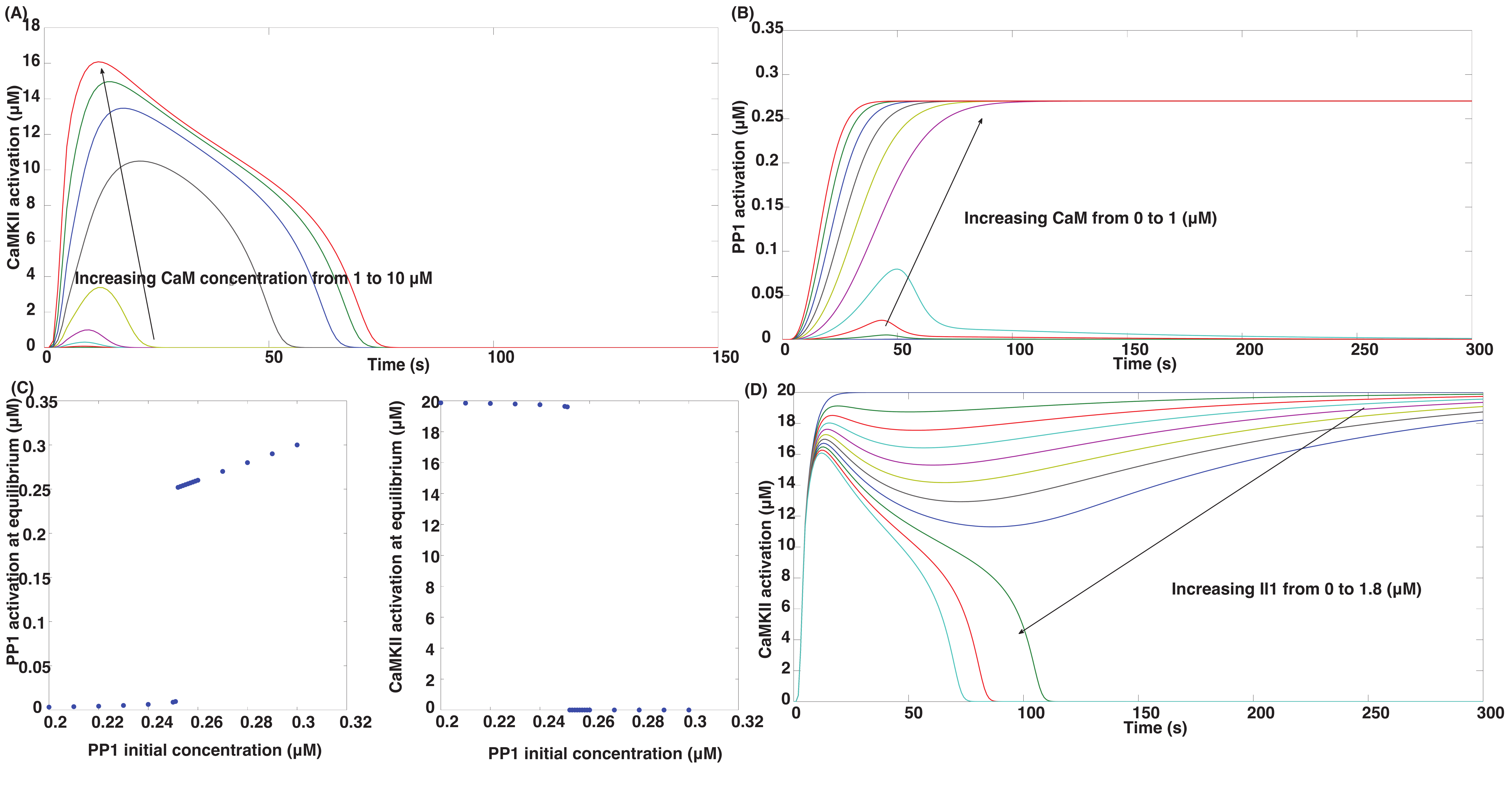}}
\caption{(A) CaMKII dynamics depends on the amount of CaM. If the concentration of CaM is low, not enough CaMKII is activation through phosphorylation. (B) PP1 dynamics is also affected by CaM. For very low amounts of CaM, PP1 activation is not enough to dephosphorylate CaMKII and it results in sustained CaMKII activation. (C) Ultrasensitivity of PP1 and CaMKII activation on PP1 initial concentration. The series of phosphatases (CaN, I1, and CaMKII) coupled with autophosphorylation of CaMKII, and autodephosphorylation of PP1 gives rise to ultrasensitive responses. (D) Increasing I1 concentration also affects CaMKII activation -- low concentration of I1 results in sustained activation of CaMKII and high concentration of I1 results in transient activation of CaMKII.}
\label{fig:figures2}
\end{figure}

\begin{figure}[!!h]
\centerline{\includegraphics[width=\textwidth,height=0.8\textheight]{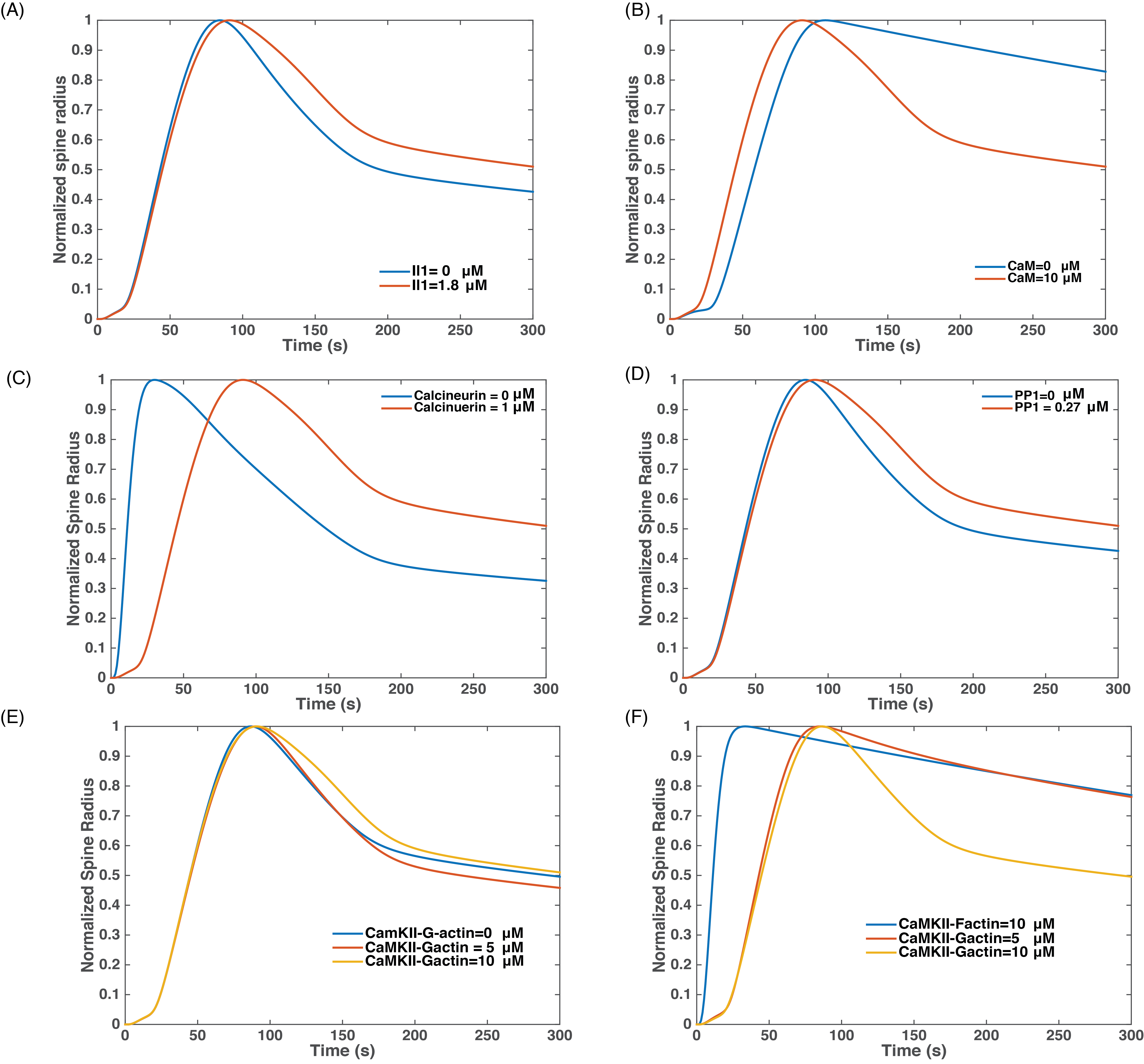}}
\caption{Effect of the components of the CaMKII module on spine radius change. Note that the normalized spine radius is shown and does not reflect the absolute change in the radius of the spine. Effect of I1 concentration (A) , PP1 (D) , and CaMKII-G-actin (E) on spine radius is small. As we explain in later sections, the dynamics downstream of these components is governed by the generation of actin barbed ends and the effect of these components is mitigated by other interactions in the network. On the other hand CaM (B), CaN (C), and CaMKII-F-actin (F) concentrations have significant impact on spine radius change. }
\label{fig:figures3}
\end{figure}

\begin{figure}[!!h]
\centerline{\includegraphics[width=\textwidth]{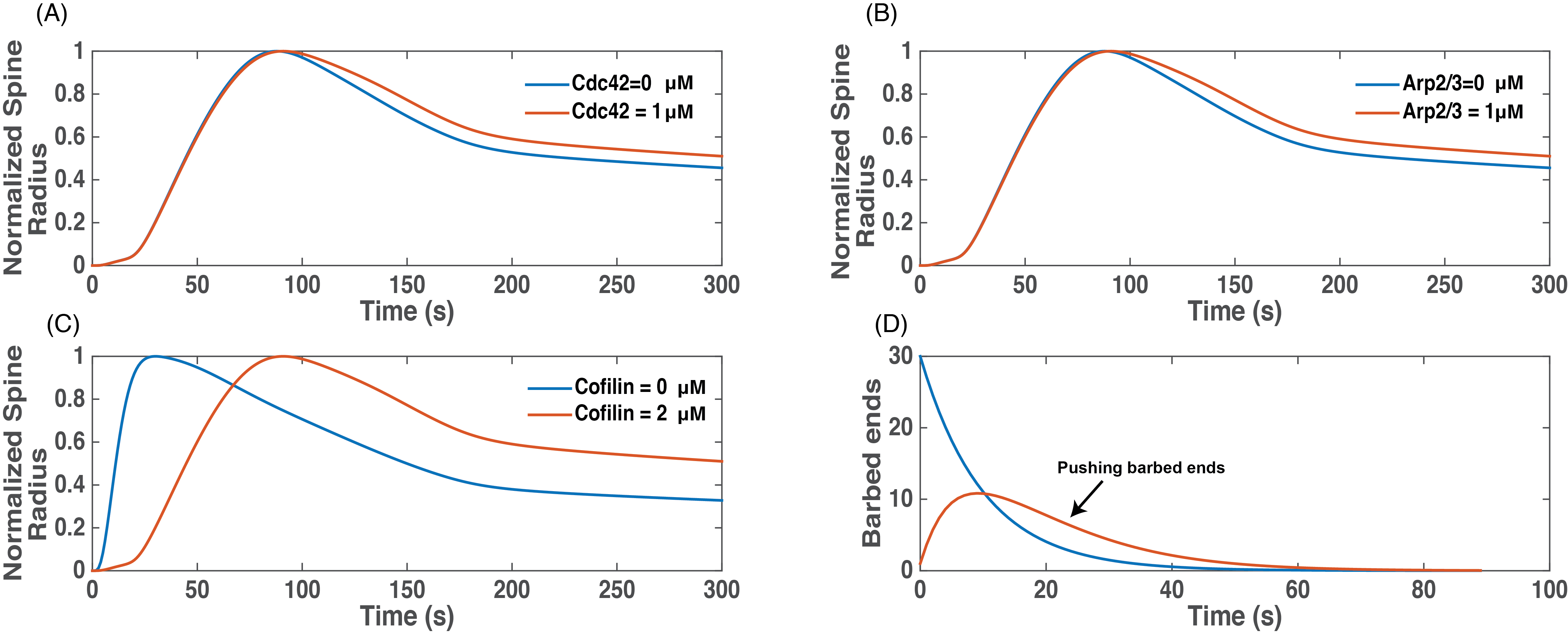}}
\caption{Effect of Cdc42 and related components on spine volume change. (A) From first observations, it seems that Cdc42 alone does not have a significant effect on the spine volume change. (B) Similarly, Arp2/3 alone does not have a strong effect on spine volume change. (C) However, cofilin has a significant effect on spine volume dynamics. This is a case where the model indicates that the complex interactions of the actin network are important for governing the dynamics of the dendritic spine. (D) When both cofilin and Arp2/3 are removed, the number of barbed ends and pushing barbed ends falls to zero very rapidly since all barbed ends get capped.}
\label{fig:figures4}
\end{figure}

\begin{figure}[!!h]
\centerline{\includegraphics[width=\textwidth, height=2in]{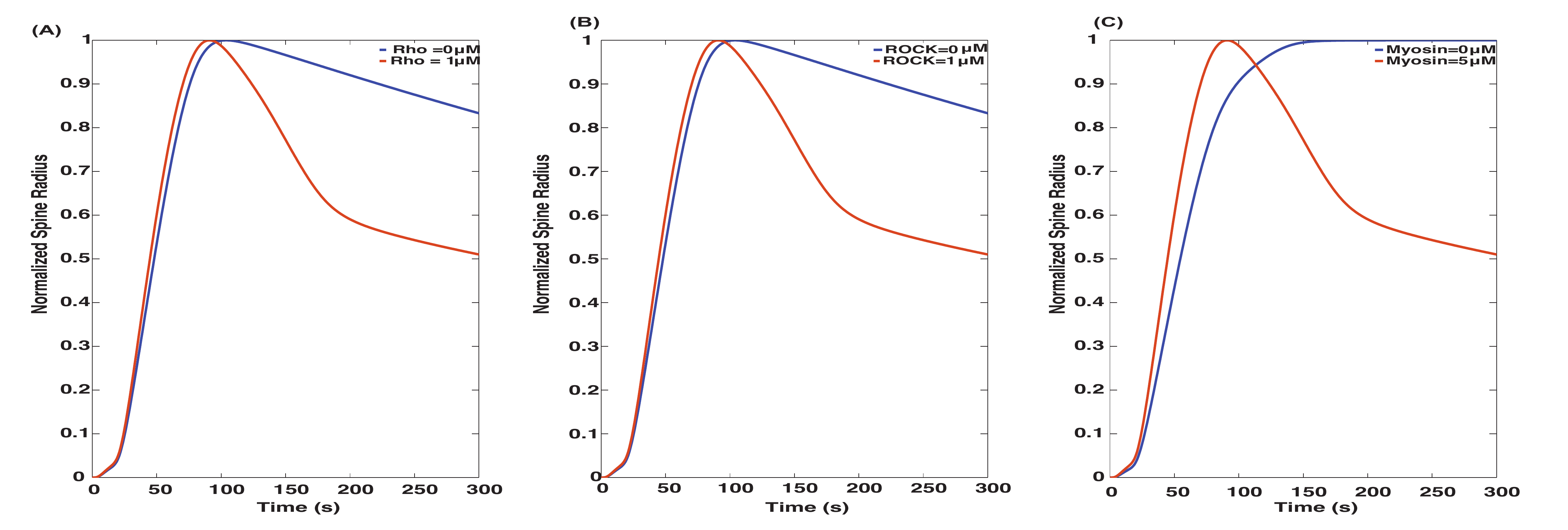}}
\caption{Effect of Rho and related components on spine volume change. (A) Rho is required for maintaining the dynamics of spine volume decrease; note that there is no effect on the increase in spine volume. (B) Rho-kinase is also important for maintaining the dynamics of spine volume decrease. (C) Myosin is the key component for governing the decrease in spine volume. Absence of myosin results in a sustained increase in spine volume.}
\label{fig:figures5}
\end{figure}

\begin{figure}[!!h]
\centerline{\includegraphics[width=\textwidth]{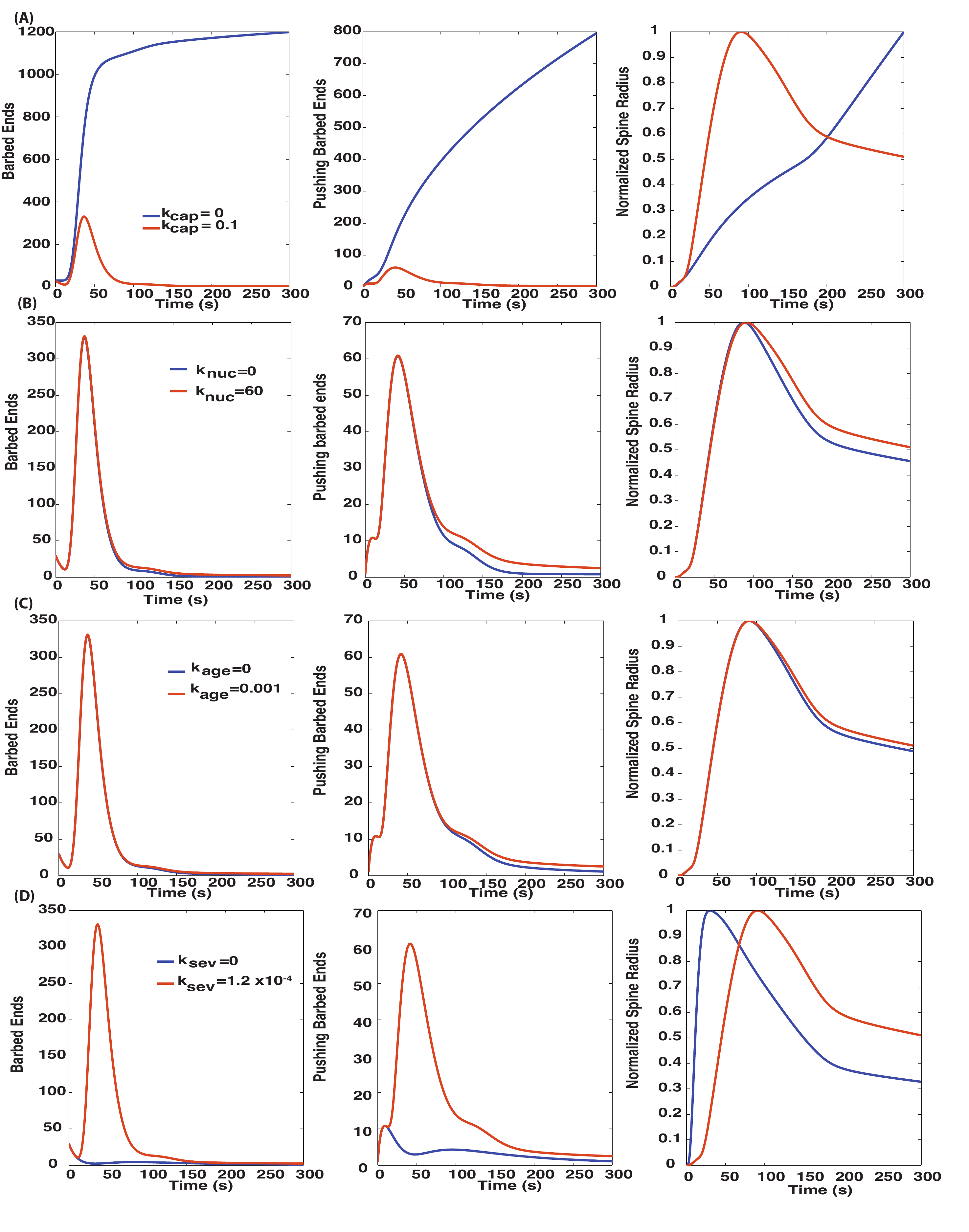}}
\caption{Effect of actin remodeling. (A) Capping rate is an important aspect of controlling spine dynamics. Setting k$_{cap}$ to zero results in a large increase in barbed ends, pushing barbed ends, and spine radius. (B) Filament nucleation rate k$_{nuc}$ and (C) filament aging k$_{age}$ do not have a significant impact on barbed ends, pushing barbed ends, and spine radius. (D) Filament severing rate k$_{sev}$ plays an important role in governing barbed end and spine dynamics. These predictions can be tested using pharmacological treatments in the laboratory.}
\label{fig:figures6}
\end{figure}

\begin{figure}[!!h]
\centerline{\includegraphics[scale=0.5]{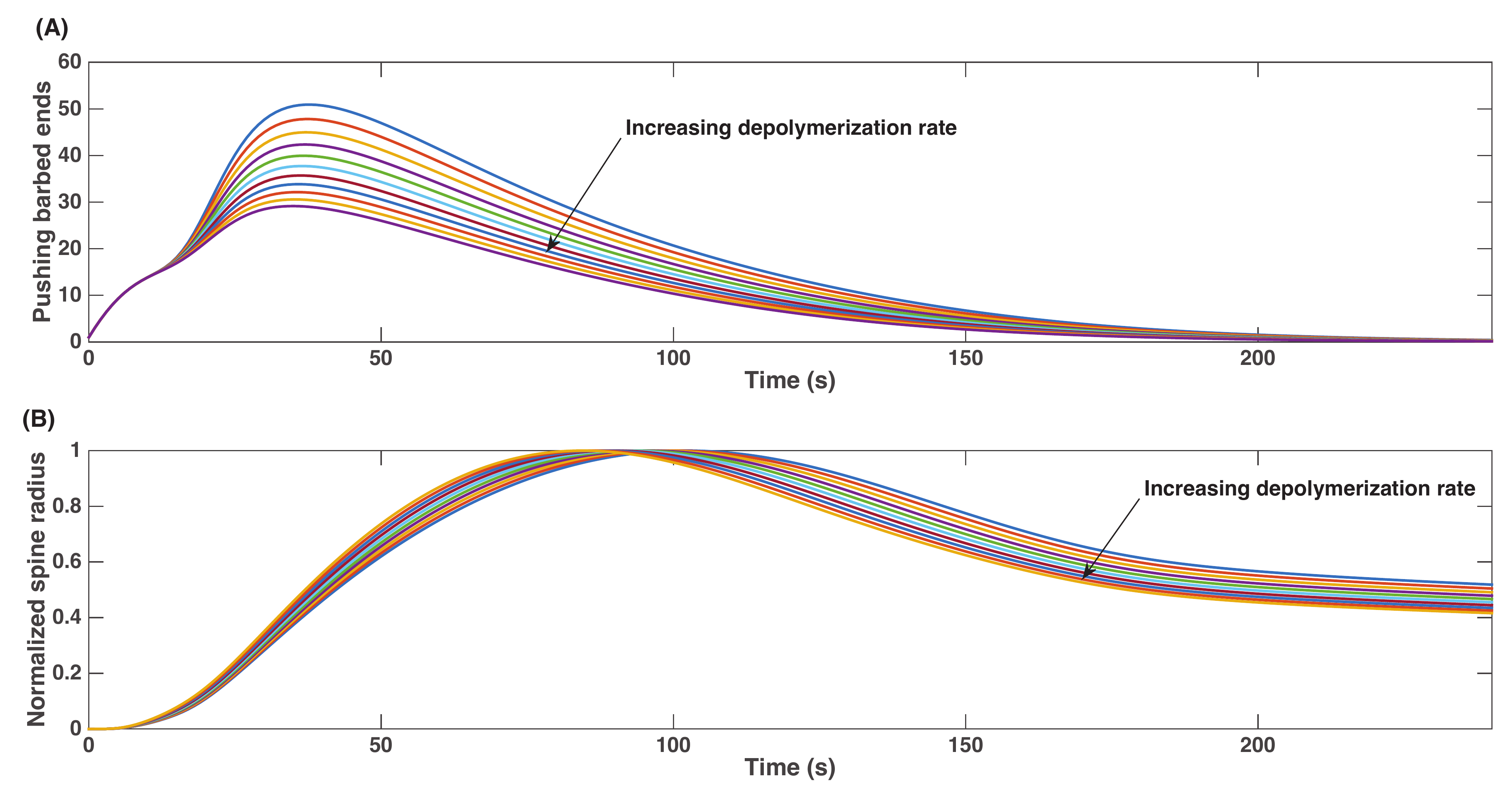}}
\caption{Role of actin depolymerization. We tested the effect of increasing depolymerization rate on pushing barbed end generation and spine radius change. (A) Increase the depolymerization rate from $0$ to $0.1$ $s^{-1}$ reduced the number of pushing barbed ends. (B) For the same values of the depolymerization rate, the spine radius did not change very much because the system was in the `coherent' regime.  }
\label{fig:figures7}
\end{figure}

\begin{figure}[!!h]
\centerline{\includegraphics[scale=0.5]{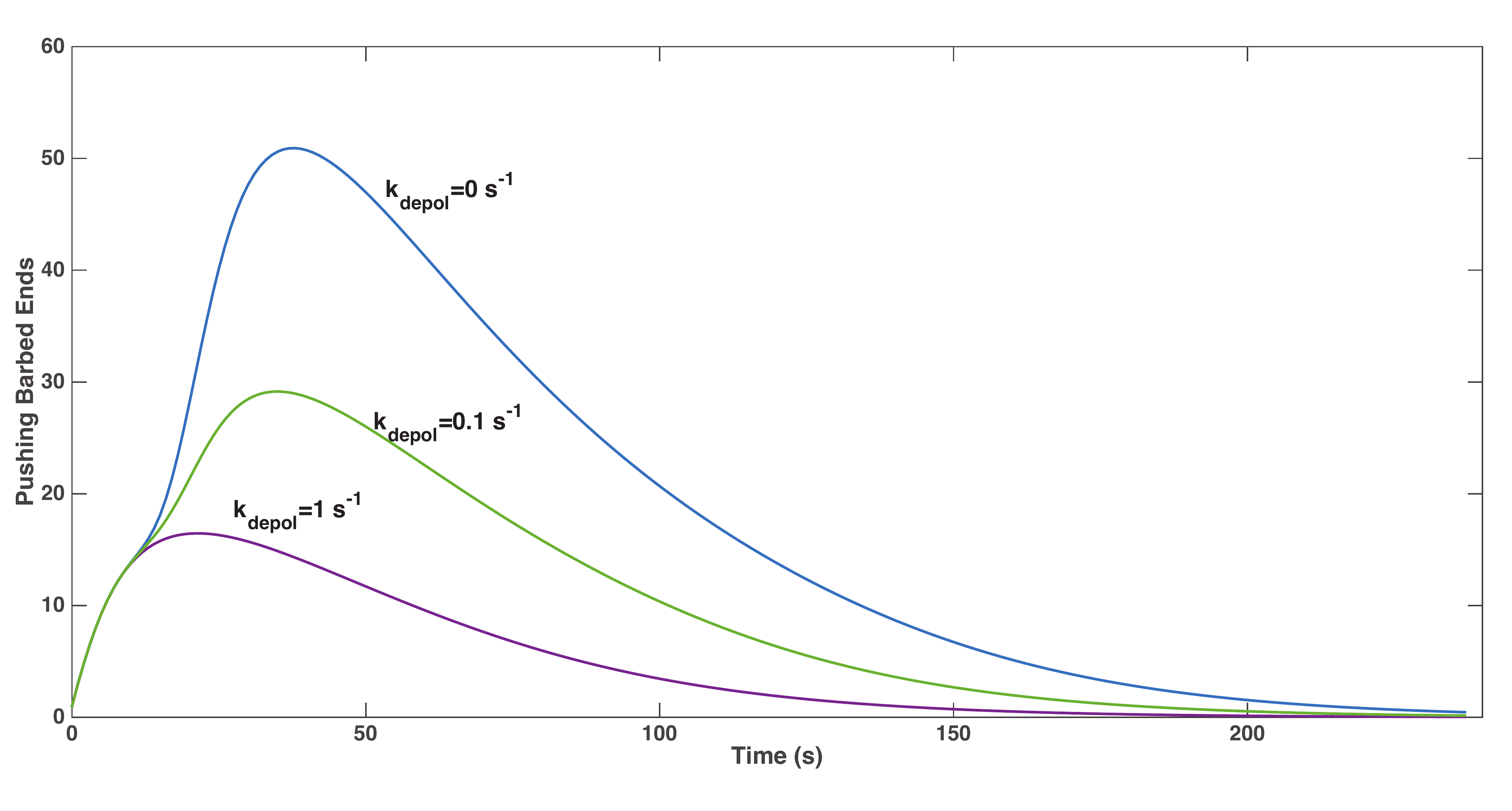}}
\caption{Effect of depolymerization rate on pushing barbed ends. Increasing the depolymerization rate decreases the number of pushing barbed ends. When the depolymerization rate is very high, there are very few pushing barbed ends available to generate a protrusive velocity.}
\label{fig:figures8}
\end{figure}

\begin{figure}[!!h]
\centerline{\includegraphics[scale=0.5]{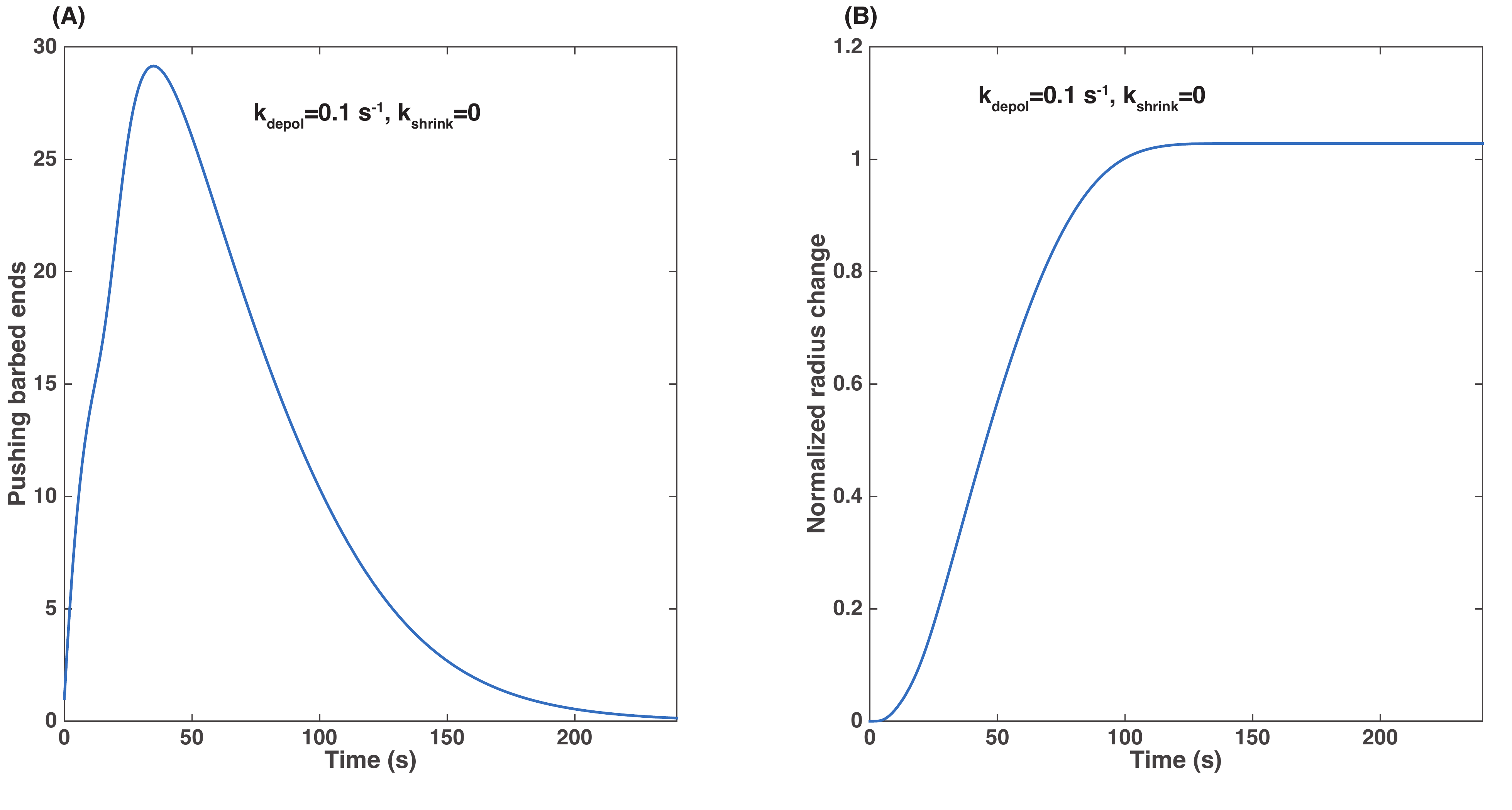}}
\caption{Actin depolymerization or myosin-contraction? If only actin depolymerization is included in the model with no myosin-mediated contraction, then (A) the pushing barbed ends are still generated and (B) increase the spine radius, but the spine radius does not decrease.}
\label{fig:figures9}
\end{figure}
\clearpage

\printbibliography

\end{document}